\pdfoutput=1
\documentclass[aps,prapplied,reprint,superscriptaddress,longbibliography,nofootinbib]{revtex4-2}
\usepackage[T1]{fontenc}
\usepackage{amsmath,amssymb,bm}
\usepackage{graphicx}
\usepackage{booktabs}
\usepackage{array}
\usepackage[colorlinks=true,linkcolor=blue,citecolor=blue,urlcolor=blue]{hyperref}
\makeatletter\let\switch@array\relax\makeatother

\newcolumntype{L}[1]{>{\raggedright\arraybackslash}p{#1}}
\newcommand{\nbar}{\bar{n}}
\newcommand{\ket}[1]{\lvert #1 \rangle}
\newcommand{\dket}[1]{\lvert \tilde{#1} \rangle}

\begin{document}
\raggedbottom

\title{Photon-stimulated decay of $\chi$-matched ancillas causes unheralded bit flips in bosonic codes}

\author{Aleksandr Gulkanov}
\affiliation{Applied Physics, Rice University, Houston, TX, USA}
\date{\today}

\begin{abstract}
Bosonic codes often read out their syndrome with a transmon ancilla whose computational levels $g$ and $f$ have equal
parity and matched dispersive shifts: a decay $f\to e$ is then heralded and leaves the code intact, while the direct
$f\to g$ is forbidden by parity. We show that the photons of the code open this decay at first order in the ancilla's
decay rate. Dressing by the code mode lets the charge noise that limits the ancilla's $T_1$ drive a photon-stimulated
Raman decay $\ket{f,n}\to\ket{g,n+1}$, at $(n+1)(\chi_{ge}/2g_{01})^2$ times the $f\to e$ rate for a smooth noise
spectrum. When the matching is set by the detuning, the decay costs $\approx\Gamma_{fe}/\alpha$ per gate and photon,
whatever the coupling strength; a deliberate mismatch or matching by a drive lowers this cost, and a qubit ancilla has no
such channel. The opened decay is unheralded and flips a cat with probability $\approx\frac12$; being first order, it
overtakes the second-order double decay $f\to e\to g$ as the ancilla improves. In a microscopic model of the Ocelot
repetition cat code, it makes the bit-flip floor of the gates grow linearly with the cat size, by a factor 1.2--1.7 from
four to ten photons, which would limit the noise bias to $\sim10^3$. The published gate data end at four photons and do
not yet test this growth.
\end{abstract}

\maketitle

\section{Introduction}
\label{sec:intro}
\enlargethispage{2\baselineskip}
Bosonic codes store a qubit in a harmonic mode and read the syndrome out through a nonlinear ancilla, usually a
transmon~\cite{Rosenblum2018Faulttolerant,Reinhold2020Errorcorrected,Eickbusch2022Fast,Sivak2023Realtime,Putterman2025Hardwareefficient}.
A single ancilla decay during the readout must not corrupt the code. A common remedy is an ancilla with two
computational levels of equal parity: the $g$ and $f$ levels of a transmon whose dispersive shifts are matched
($\chi$-matching), so that a decay $f\to e$ shifts the code by a known amount and can be
heralded~\cite{Rosenblum2018Faulttolerant,Reinhold2020Errorcorrected}, while the direct $f\to g$ is forbidden: ``selection
rules forbid direct $f$ to $g$ transitions''~\cite{Rosenblum2018Faulttolerant}. The most demanding use is in repetition cat
codes, whose cats suppress bit flips exponentially in their photon number $\nbar$, so that a repetition code against the
remaining phase flips suffices~\cite{Guillaud2019Repetition,Puri2020Biaspreserving,Chamberland2022Building,LeRegent2023Highperformance,Gouzien2023Performance}.
In the repetition-cat-code memory of AWS (``Ocelot'')~\cite{Putterman2025Hardwareefficient} the bit flips come mainly from
the $(g,f)$ transmon ancillas that extract the syndrome: in repeated gate cycles the bit-flip probability per cycle is
$(3.5\pm0.4)\times10^{-3}$ at $\nbar=2$ with the ancilla in $\ket g+\ket f$, and at $\nbar=4$ the bit-flip time is about
1.5~ms with the ancilla in $\ket g+\ket f$ but exceeds 5~ms with it in $\ket g$.

\begin{figure*}
\includegraphics[width=\textwidth]{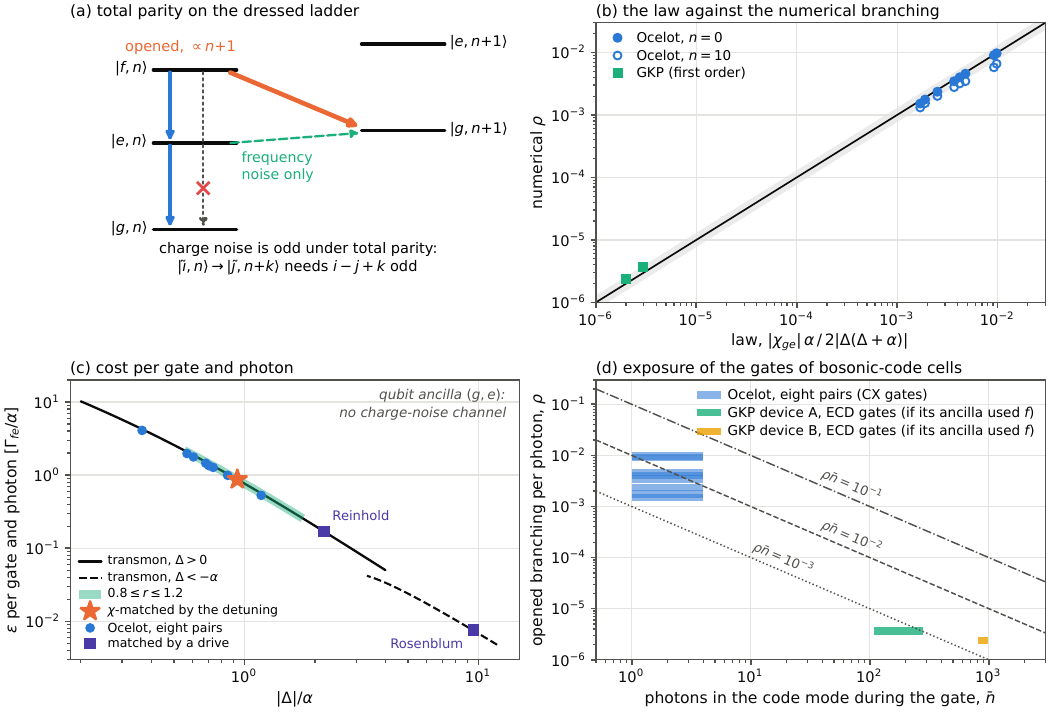}
\caption{The opened channel. (a) Dressed ladder of a transmon ancilla and a bosonic mode slightly above its $g$--$e$
frequency ($\Delta>0$; schematic). Charge noise is odd under total parity: it connects $\ket{\tilde i,n}$ and
$\ket{\tilde j,n+k}$ only if $i-j+k$ is odd, so $f\to g$ needs a photon, and the one-photon $e\to g$ is left to
frequency noise. (b) Opened branching per photon, numerical against the law, Eq.~\eqref{eq:law}: the Ocelot pairs at
$n=0$ (filled) and $n=10$ (open) and the GKP devices at first order (App.~\ref{app:gkp}); shaded, within a factor 1.3
(Table~\ref{tab:law}). (c) Cost per gate and photon, Eq.~\eqref{eq:pergate}, against the detuning for a directly coupled
transmon ($E_J/E_C=95$; mode above, solid, or below the $e$--$f$ resonance, dashed; highlighted for $0.8\le r\le1.2$), with
the matching point (star), the Ocelot pairs (circles) and the drive-matched ancillas of
Refs.~\cite{Reinhold2020Errorcorrected,Rosenblum2018Faulttolerant} (squares). (d) $\rho$ at the photon numbers of the Ocelot
CX gates and of the large ECD displacements of the GKP devices (as if their ancilla used $f$); lines, constant
$\rho\nbar$.}
\label{fig:channel}
\end{figure*}

These errors have so far been described with an ancilla decay that does not depend on the photon number. The device's own
simulations of its gates take the measured lifetimes of $e$ and $f$, divided by a factor fitted to the gate
errors~\cite{Putterman2025Hardwareefficient}; architecture models treat the ancilla floor as independent of
$\nbar$~\cite{Hann2025Hybrid}; and the experiment names the prediction of the gate errors from independently measured
quantities an important future direction, and a direct $f\to g$ decay, possibly mediated by defects, among the error
mechanisms still to be investigated~\cite{Putterman2025Hardwareefficient}.

We ask whether $\chi$-matching still protects the code at first order in the ancilla's decay when the code mode holds
photons, and find that it leaves one channel open. Charge noise, which sets the $T_1$ of a transmon, is odd under total
parity and cannot drive $f\to g$ without a photon, but dressing by the code mode opens a photon-stimulated Raman decay
$\ket{f,n}\to\ket{g,n+1}$ (Sec.~\ref{sec:parity}); at $\chi$-matching by the detuning it costs $\approx\Gamma_{fe}/\alpha$
per gate and photon, whatever the coupling (Sec.~\ref{sec:cost}). Unheralded and first order in the decay rate, it leads
the silent faults once its branching exceeds the probability that the ancilla decays from $e$ during the gates
(Sec.~\ref{sec:conversion}), and in repetition cat codes it makes the bit-flip floor of the gates grow as $c_0+c_1\nbar$,
which would cap the noise bias and gives the logical error a true minimum in $\nbar$ (Sec.~\ref{sec:consequences}).

This work adds the channel's action on the ancilla of a bosonic code during every gate: its law in dispersive parameters,
the coupling-independent cost per gate that follows from $\chi$-matching, its place among the faults of syndrome
extraction, and its consequences for repetition cat codes; the ingredients, total parity and the cavity-assisted
$f0$--$g1$ Raman element, are known (Sec.~\ref{sec:earlier}). The photons that protect a
bosonic code thus act on its ancilla like readout photons on a
qubit~\cite{Sank2016MeasurementInduced,Dumas2024MeasurementInduced,Connolly2025Full,Dai2026Characterization}, during every
gate, and total parity decides which of their channels open.

We confront the channel with the published Ocelot data through a microscopic model of its cells, calibrated on static
spectroscopy, never on error rates (Sec.~\ref{sec:ocelot}). The model singles out the cell whose dispersive shifts are
mismatched, but not the order of the other seven; the measured level is about four times the model's, and measured
on-gate decay accounts for half of that factor. The data end at four photons, before the floor would visibly grow, so the growth is a prediction
(Sec.~\ref{sec:discussion}).

\section{The opened channel}
\label{sec:parity}

\subsection{Selection rule}

Consider a transmon ancilla~\cite{Koch2007Chargeinsensitive} with charge operator $\hat n$ coupled to a bosonic mode,
\begin{equation}
H=H_t(\hat n,\hat\varphi)+\omega\,a^\dagger a+g\,\hat n\,(a+a^\dagger).
\label{eq:H}
\end{equation}
At offset charge $n_g=0$, and up to the exponentially small charge dispersion otherwise, $H_t$ is invariant under the
transmon parity $\Pi_t$ ($\varphi\to-\varphi$), which makes $\hat n$ odd (at $n_g=\frac12$ the odd operator is
$\hat n-\frac12$, and the remainder only displaces the mode); $H$ then conserves
$\Pi=\Pi_t(-1)^{a^\dagger a}$~\cite{Wallraff2007Sideband,Beaudoin2011Dissipation,Cohen2023Reminiscence}. Charge
noise---dielectric loss and the dipoles of two-level defects, which set the $T_1$ of
transmons~\cite{Martinis2005Decoherence,Muller2019Towards}---couples through $\hat n$ and is $\Pi$-odd: with
$\Pi_t\ket i=(-1)^i\ket i$ it connects the dressed states $\ket{\tilde i,n}$ and $\ket{\tilde j,n+k}$ only if $i-j+k$ is
odd, and a $\Pi$-even operator such as frequency noise only if $i-j+k$ is even [Fig.~\ref{fig:channel}(a)].

For a qubit ancilla $(g,e)$, charge noise therefore drives decay only with an even change of the photon number; the
one-photon decay $\ket{\tilde e,n}\to\ket{\tilde g,n+1}$ needs $\Pi$-even noise, the dressed dephasing of dispersive
readout~\cite{Boissonneault2008Nonlinear,Boissonneault2009Dispersive,Slichter2012MeasurementInduced,Kahn2026Qubit}. For the
equal-parity pair $(g,f)$ the photon-free decay is forbidden, but charge noise drives $\ket{\tilde f,n}\to\ket{\tilde
g,n+1}$, with one photon into the mode and a quantum $\omega_{gf}-\omega$ into the bath; in a multilevel ancilla it opens
$j\to j-2$ for every $j\ge2$. The other photon-number-dependent charge channels of a $(g,f)$ ancilla, the two-photon
heating $\ket{\tilde g,n}\to\ket{\tilde e,n-2}$ and $\ket{\tilde f,n}\to\ket{\tilde e,n\pm2}$, are second order and end in
$e$ (the heating $\lesssim2\times10^{-5}$ per CX$^2$ in the Ocelot pairs at $\nbar\le4$). Here $\alpha>0$ is the magnitude of
the anharmonicity, $\Delta=\omega-\omega_{ge}$, $\chi_{ge}$ and $\chi_{gf}$ are the shifts of $\omega_{ge}$ and $\omega_{gf}$
per photon (twice the $\chi$ of Ref.~\cite{Koch2007Chargeinsensitive}), frequencies are angular, and $\nbar=|\alpha|^2$ is
the mean photon number of a cat; where $|\alpha|^2$ appears, $\alpha$ is the cat amplitude, elsewhere the anharmonicity. We
illustrate the results with the eight storage--ancilla cells of Ocelot, in which a flux-tunable coupler turns on the
storage--ancilla interaction during the conditional rotations, at detunings $\Delta/2\pi=70$--190~MHz
(Sec.~\ref{sec:ocelot}).

With a tunable coupler the conserved operator becomes $\mathcal P=(-1)^{k+c+n}$ (ancilla level $k$, coupler level $c$,
photon number $n$); the charges and $a+a^\dagger$ are $\mathcal P$-odd, and every charge element between dressed states of
the full cell with equal parity of $k+n$ vanishes, numerically to $10^{-30}$. Charge noise therefore cannot drive
$\ket{\tilde g,n}\to\ket{\tilde e,n-1}$, $\ket{\tilde f,n}\to\ket{\tilde g,n}$, $\ket{\tilde f,n}\to\ket{\tilde e,n+1}$ or
$\ket{\tilde f,n}\to\ket{\tilde g,n+2}$. Offset charge, junction asymmetry and the flux ramps break this parity only
weakly (App.~\ref{app:model}).

\subsection{The opened element and its law}
\label{sec:law}

To first order in the coupling, the opened element between $\ket{\tilde f,n}$ and $\ket{\tilde g,n+1}$ is
\begin{equation}
\begin{split}
M_{fg}(n)={}&g\sqrt{n+1}\sum_kn_{gk}n_{kf}\\
&\times\left(\frac1{E_{fk}-\omega}+\frac1{E_{gk}+\omega}\right),
\end{split}
\label{eq:M}
\end{equation}
$E_{ik}=E_i-E_k$: two interfering paths through the intermediate levels. It is the element of the cavity-assisted
$f0$--$g1$ Raman process~\cite{Pechal2014Microwave,Zeytinoglu2015Microwave}, with the drive replaced by the bath operator
and stimulated by the $n$ photons of the mode; the rate $\Gamma_{fg}(n)=|M_{fg}|^2S_{nn}(\omega_{gf}-\omega)$ grows as
$n+1$. The absorption branch $\ket{\tilde f,n}\to\ket{\tilde g,n-1}$ nearly cancels at first order ($\approx5\%$ of the
emission rate in GKP device A) and is neglected. In the full Ocelot cell the squared element grows linearly at small
$n$ and more slowly near the critical photon number; the coupler-charge channel is 12--14 times weaker
(App.~\ref{app:spectra}).

For a transmon only $k=e$ matters; with $n_{ef}\simeq\sqrt2n_{ge}$ the two paths give $M_{fg}\simeq
g_{01}\sqrt{n+1}\,n_{ef}\,\alpha/[\Delta(\Delta+\alpha)]$, and with $\chi_{ge}\simeq-2g_{01}^2\alpha/[\Delta(\Delta+\alpha)]$
the branching per photon relative to the $f\to e$ decay in a flat bath becomes
\begin{equation}
\rho\equiv\frac{\Gamma_{fg}}{(n+1)\Gamma_{fe}}\approx\frac{|\chi_{ge}|\alpha}{2|\Delta(\Delta+\alpha)|} ,
\label{eq:law}
\end{equation}
independent of the coupling; with $g_{01}=g|n_{ge}|$ it equals $(\chi_{ge}/2g_{01})^2$. Against the exact branching of the
eight Ocelot pairs (dressed ladders of the cell model, relative to the dressed $f$--$e$ element at $n=0$) and the
first-order branching of the two GKP devices, the law holds to 0.90--0.99 and 1.21--1.24 at $n=0$
[Fig.~\ref{fig:channel}(b)]; the GKP excess comes from the counter-rotating terms that the two-path form omits. At $n=10$
the Ocelot $\rho$ has fallen to 0.64--0.88 of its $n=0$ value; for $n\le2$ the law holds within 13\%.

Equation~\eqref{eq:law} fixes the dressing; the rate is
\begin{equation}
\Gamma_{fg}(n)=\sigma\,(n+1)\,\rho\,\Gamma_{fe},\qquad \sigma\equiv\frac{S_{nn}(\omega_{gf}-\omega)}{S_{nn}(\omega_{ef})},
\label{eq:sigma}
\end{equation}
where $\sigma$ compares the charge noise at the released frequency $\omega_{gf}-\omega=\omega_{ef}-\Delta$ with that at
$\omega_{ef}$, only 70--190~MHz away in Ocelot. A smooth bath has $\sigma=1$, which we assume unless stated otherwise;
two-level defects make $\sigma$ device-specific, the more so as the dispersive shift sweeps the released quantum
(4.8--5.0~GHz in Ocelot) through a band of tens of MHz (Sec.~\ref{sec:ocelot} and App.~\ref{app:tls}).

\subsection{Relation to earlier work}
\label{sec:earlier}

The total-parity rule is standard in circuit QED~\cite{Wallraff2007Sideband,Beaudoin2011Dissipation,Cohen2023Reminiscence}.
The opened element is that of the cavity-assisted $f0$--$g1$ Raman process, driven coherently to emit single
photons~\cite{Pechal2014Microwave,Zeytinoglu2015Microwave}, whose driven sideband shows its bosonic enhancement up to
$n=14$~\cite{Huang2026Fast}; its environment-driven counterpart has been observed in the other direction, as readout-induced
transitions $\ket g\to\ket f$ at rates proportional to the photon number, which a parameter-free theory reproduces to
$\approx25\%$~\cite{Connolly2025Full}. The operator appears, and is dropped, in perturbative theories
of readout-induced decay~\cite{Malekakhlagh2020Lifetime,Petrescu2020Lifetime}, and Floquet--Markov
treatments~\cite{Hone2009Statistical} of driven transmons~\cite{Cohen2023Reminiscence} contain it as a sideband. Direct $\ket2\to\ket0$ decays measured
without photons are orders of magnitude faster than the opened channel and come from other
processes~\cite{Peterer2015Coherence}. We found no earlier evaluation of the channel for bosonic-code ancillas.

\section{Cost per gate and which ancillas are exposed}
\label{sec:cost}

\subsection{Cost per gate and \texorpdfstring{$\chi$}{chi}-matching}

A conditional $\pi$ rotation lasts $T=\pi/|\chi_{gf}|$, so the opened probability per gate and photon is
\begin{equation}
\varepsilon\equiv\rho\,\Gamma_{fe}T\approx\frac{\pi r}{2}\,\frac{\Gamma_{fe}\,\alpha}{|\Delta(\Delta+\alpha)|}=\frac{\pi r}{2x(1+x)}\,\frac{\Gamma_{fe}}{\alpha},
\label{eq:pergate}
\end{equation}
times $\sigma$ for a structured bath, with $r=\chi_{ge}/\chi_{gf}$ and $x=\Delta/\alpha$: at a given detuning neither the
coupling nor the gate speed enters [Fig.~\ref{fig:channel}(c)]. $\chi$-matching ($r=1$), which makes a single $f\to e$
decay harmless~\cite{Rosenblum2018Faulttolerant}, fixes $x$ when it is set by the static shifts: a directly coupled
transmon matches at $x=0.900$, 0.932 and 0.949 for $E_J/E_C=50$, 95 and 150, a larger detuning breaks the matching
($r=0.69$ at $x=3$), and with the mode below the $e$--$f$ resonance ($x<-1$) the static shifts cannot be matched ($r<0.45$
for mode frequencies above 1~GHz). In Ocelot the coupler moves the matching point to $x\approx0.80$, and the model
reproduces the measured $r$ of all eight pairs to $\le0.011$ (Table~\ref{tab:chimatch} in App.~\ref{app:model}). At exact
matching by the detuning the cost is therefore $(0.85$--$1.1)\,\Gamma_{fe}/\alpha$, $\approx3\times10^{-5}$ per gate and
photon for $\alpha/2\pi=200$~MHz and $T_1^{fe}=25~\mu$s, whatever the coupling or the gate speed; the mismatched pair
A4--S4 ($r=1.30$, $x=0.37$) pays $1.35\times10^{-4}$ (at its $\alpha/2\pi=192$~MHz). A deliberate mismatch lowers the cost, 2.5 times at $x=1.5$
($r=0.85$), at the price of residual rotations after $f\to e$ decays, which the re-stabilization of the cat corrects for
$0.8\lesssim r\lesssim1.2$~\cite{Putterman2025Hardwareefficient}; so do a longer $T_1^{fe}$, a larger $\alpha$, and
matching at larger detuning by a drive: the drive-matched ancillas of
Refs.~\cite{Reinhold2020Errorcorrected,Rosenblum2018Faulttolerant} pay $4.2\times10^{-6}$ and $2.5\times10^{-7}$ per gate
and photon, five and a hundred times below the same transmons matched by the detuning (App.~\ref{app:model}). The cost
$\approx\Gamma_{fe}/\alpha$ thus applies to ancillas matched by their static shifts, as in Ocelot, which matches natively by
the detuning~\cite{Putterman2025Hardwareefficient}.

\subsection{Which ancillas are exposed}
\label{sec:exposure}

\looseness=1 
The law sorts bosonic-code cells by their exposure [Fig.~\ref{fig:channel}(d)]. Ocelot's CX gates operate at
$\Delta=0.4$--$1.2\alpha$ and reach $\rho=1.5$--$9.7\times10^{-3}$ per photon, so at its 1--4 photons the opened channel is
0.3--4\% of the $f\to e$ decay. GKP qubits stabilized with echoed conditional displacements
(ECD)~\cite{Eickbusch2022Fast,Sivak2023Realtime} drive their cavity to hundreds of photons but are detuned by about seven
anharmonicities, so they would reach only $\rho\approx3\times10^{-6}$ if their ancilla used $f$; their qubit ancilla has
no one-photon charge channel at all, which a Floquet--Markov treatment of two published devices confirms beyond the
dispersive regime (App.~\ref{app:gkp}). Multilevel ancillas acquire one-photon jumps $j\to j-2$, which concerns designs whose
protection assumes single-level steps, such as the $0$--$n$ Fock qubits of Ref.~\cite{Direkci2026Dissipatively}, not
evaluated here.

\begin{figure*}
\includegraphics[width=\textwidth]{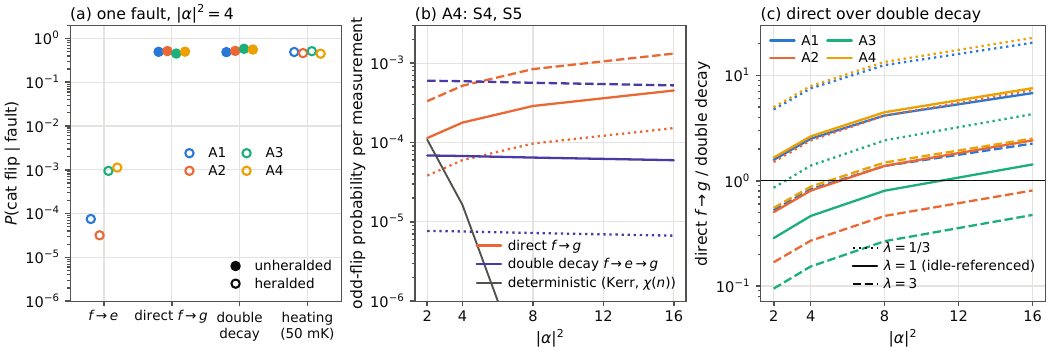}
\caption{Conversion of ancilla faults into cat errors in the Ocelot syndrome cycle (model). (a) Probability that one fault
of each class leaves an odd number of cat flips in a stabilizer measurement at $|\alpha|^2=4$ (idle-referenced bath;
heating from a 50-mK variant); filled, unheralded; open, heralded. (b) Direct $f\to g$ and double decay $f\to e\to g$ in
the logical-relevant error of the A4 stabilizer for $\lambda=1/3$ (dotted), 1 (solid) and 3 (dashed), with the
deterministic part of the gates. (c) Their ratio for the four stabilizers and the same baths.}
\label{fig:conversion}
\end{figure*}

\section{Why the channel matters: conversion into cat errors}
\label{sec:conversion}

\looseness=1 
In the repetition cat code each ancilla, prepared in $(\ket g+\ket f)/\sqrt2$, performs a CX with each of its two cats, is
unprepared, read out ($e$ is reported as an erasure herald) and reset. We propagate both ancilla branches through the two
gates with the dressed-ladder model of the Ocelot cells, re-stabilize each cat with the asymptotic map of the two-photon
dissipation, and split the evolution into fault sequences with at most one jump of each class ($f\to e$, $e\to g$, direct
$f\to g$, heating) by a multi-class first-order Dyson expansion that closes on the full evolution
(App.~\ref{app:conversion}). Each sequence gets its probability per stabilizer measurement, its probability of an odd number
of cat flips (the logical-relevant error, since the repetition code does not correct bit flips) and its herald probability;
the decay rates at the operating point are idle-referenced unless scaled by $\lambda$.

\textit{One fault.}---The conversion separates the faults sharply [Fig.~\ref{fig:conversion}(a)]. A single $f\to e$ decay,
by far the most probable fault (1.4--2.0\% per measurement at $|\alpha|^2=4$), is heralded and flips a cat with
probability $\lesssim10^{-3}$: the protection of $\chi$-matching. The opened decay leaves the ancilla in $g$, silent, and
stops the conditional rotation at a random angle, which the re-stabilization turns into a flip with probability
0.45--0.52. The double decay $f\to e\to g$ is equally silent and converts as often (0.49--0.58): an
$e\to g$ during the first CX makes the second miss entirely. Heating out of $g$ converts as often but ends in $e$ and is
heralded. Among the single charge-noise faults of a $\chi$-matched ancilla, the opened decay is therefore the only one
both likely to flip a cat and unheralded; parity-even processes, such as a defect that modulates the Josephson energy, and
coupler excitations can act alike (Sec.~\ref{sec:ocelot} and App.~\ref{app:spectra}). Single $f\to e$ decays matter only
for the mismatched pairs at low $|\alpha|^2$, where they leave a residual rotation ($0.14$--$0.16\times10^{-3}$ per
measurement for A3 and A4 at $|\alpha|^2=2$).

\looseness=1 
\textit{Weights.}---Weighted by their probabilities [Fig.~\ref{fig:conversion}(b) for A4, Fig.~\ref{fig:faultconversion}
for all four], at $|\alpha|^2=4$ the direct $f\to g$ makes 29--66\% of the logical-relevant error of the four stabilizers,
the double decay 24--62\% and the deterministic Kerr shear of the unstabilized gate $\le7\%$, of a total of
$0.23$--$0.30\times10^{-3}$ per measurement; heating at 50~mK would raise the total by 9--20\%. At $|\alpha|^2=16$ the direct $f\to g$ makes 58--85\% and carries the entire photon-number-linear term,
$0.010$--$0.024\times10^{-3}$ per photon and measurement, while the double decay stays flat ($0.06$--$0.15\times10^{-3}$).

\looseness=1 
\textit{Order in the decay rate.}---The opened decay is first order in the ancilla's decay rates, the double decay second
order: rates three times larger multiply them by 2.9 and 8.6--8.8, rates three times smaller divide them by 3.0 and 8.9
[Fig.~\ref{fig:conversion}(b), (c)]. Their ratio follows an estimate in measurable quantities, $(\nbar+1)\bar\rho/(\Gamma_{eg}T_{ab})$
times 1.1--1.8 over the four stabilizers, the three baths and $|\alpha|^2=4$--16, with $\bar\rho$ the mean branching of the
stabilizer's two pairs at $n=0$, $\Gamma_{eg}=\lambda/T_1^{ge}$ and $T_{ab}$ the summed length of its two CX gates
(App.~\ref{app:conversion}): the opened decay leads the silent faults once its photon-stimulated branching exceeds the
probability that the ancilla decays from $e$ during the two gates. With $\lambda=3$, the factor that matches the gate data
(Sec.~\ref{sec:ocelot}), the direct channel takes over from the double decay between $|\alpha|^2=4$ and 8 for A1 and A4,
and not by 16 for A2 and A3. With an
ancilla three times longer-lived ($\lambda=1/3$) the direct channel exceeds the double decay for all four stabilizers at
$|\alpha|^2\ge4$, by 1.4--7.9 times at $|\alpha|^2=4$ and up to 23 times at 16, and makes 46--70\% of the logical-relevant
error at $|\alpha|^2=4$ and 79--92\% at 16.
The opened decay is thus the charge-noise error that $\chi$-matching leaves at first order: its weight grows with the
photon number and as the ancilla improves.

\textit{The floor.}---Per cat, the gate part of the bit-flip probability per cycle therefore approaches $c_0+c_1\nbar$
(Sec.~\ref{sec:consequences}), with the slope $c_1$ set by the opened channel. Monitoring the photons that the buffer
emits while it re-stabilizes a rotated cat~\cite{Ferrari2026Bit} could herald at most about half of these errors: a gate
skipped after an early decay to $g$ leaves its cat in the code space and no trace, and the threshold needed against false
heralds from the deterministic shear discards part of the rest (App.~\ref{app:herald}).

\begin{figure*}
\includegraphics[width=\textwidth]{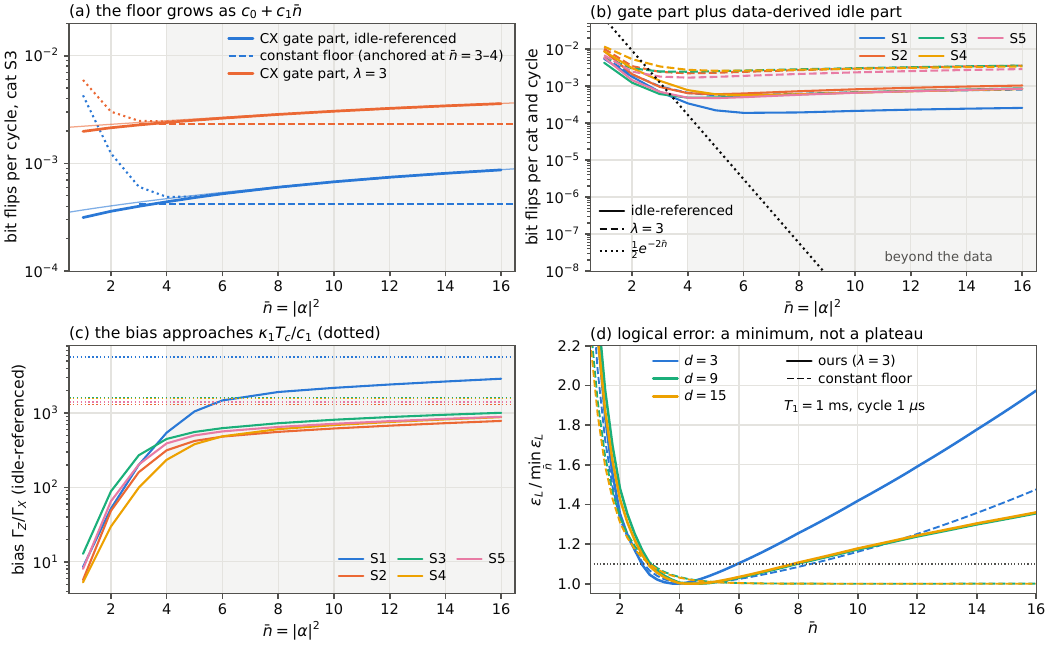}
\caption{Consequences for the Ocelot repetition cat code (model; shaded: beyond the published gate data, $|\alpha|^2>4$).
(a) CX gate part of the bit flips per 2.8-$\mu$s cycle of cat S3 ($\lambda=1$ and 3), with fits $c_0+c_1\nbar$ (thin), a
constant floor anchored on the model at $\nbar=3$--4 (dashed) and the total with the data-derived idle part (dotted). (b)
Bit flips per cat and cycle (gate plus idle part) for the five cats, idle-referenced (solid) and $\lambda=3$ (dashed);
dotted, $\frac12e^{-2\nbar}$. (c) Noise bias $\Gamma_Z/\Gamma_X$ per cat (idle-referenced) and the ceilings $\kappa_1T_c/c_1$
(dotted), 2.7--2.9 times lower at $\lambda=3$. (d) Logical error per cycle relative to its minimum for $d=3$, 9 and 15,
storage $T_1=1$~ms and a 1-$\mu$s cycle: ours ($\lambda=3$, solid) and the constant floor (dashed); dotted, 10\% above the
minimum.}
\label{fig:consequences}
\end{figure*}

\section{Consequences for repetition cat codes}
\label{sec:consequences}

We combine the stabilizer tables up to $|\alpha|^2=16$ with the idle bit flips derived from the measured data and a Stim
table of the phase-flip memory (App.~\ref{app:projection}), and compare with the phenomenology of
Refs.~\cite{Putterman2025Hardwareefficient,Hann2025Hybrid}, which replaces our gate part by an $|\alpha|^2$-independent CX
term; the idle part is common, so the comparison isolates the gates. For the growth of the floor and the bias we anchor
that term on our own gate part at $|\alpha|^2=3$--4, so that only the functional form differs; for the logical error and
$\nbar^*$ we anchor it on the measured gate rates. Beyond $|\alpha|^2=4$ every statement is a model prediction, and the
numbers span $\lambda=1$ to 3.

\textit{A growing floor.}---The gate part grows linearly [Fig.~\ref{fig:consequences}(a)], with
$c_1=0.8$--$3.9\times10^{-5}$ per photon for the five cats at $\lambda=1$ and 2.7--2.9 times more at $\lambda=3$, as for a
first-order channel: 1.2--1.7-fold between $|\alpha|^2=4$ and 10, where a constant floor stays flat, and 1.4--2.3 times that
floor at $|\alpha|^2=16$. The exponential suppression of the idle bit flips meets this floor at $\nbar\lesssim4$--6 (an
upper estimate, since the extrapolated idle part falls more slowly than the asymptotic $e^{-2\nbar}$)
[Fig.~\ref{fig:consequences}(b)].

\textit{A ceiling on the bias.}---Because the phase flips also grow linearly with $\nbar$, the noise bias
$\Gamma_Z/\Gamma_X$ approaches a ceiling $\kappa_1T_c/c_1$ (storage loss rate $\kappa_1$, cycle time $T_c$)
[Fig.~\ref{fig:consequences}(c)]: $0.5$--$5.7\times10^3$ for the five cats, the upper values at $\lambda=1$. It is approached
slowly, since $c_0$ dominates up to $|\alpha|^2\sim c_0/c_1\approx10$--16 ($\lambda=1$; 15--27 at $\lambda=3$): the bias
reaches 235--545 at $|\alpha|^2=4$ and 785--2900 at 16 ($\lambda=1$). An opened channel as strong as the direct-decay data of ancilla A4 indicate would lower the
ceilings of its two cats further (App.~\ref{app:tls}).

\textit{An optimal cat size.}---A repetition code does not correct bit flips, so its logical bit-flip probability is $d$
times the per-cat value $Ae^{-\gamma\nbar}+c_0+c_1\nbar$, which is minimal at $\nbar^*=\gamma^{-1}\ln(A\gamma/c_1)$. In our
model the logical error therefore has a true minimum [Fig.~\ref{fig:consequences}(d)], at $\nbar\approx4$--5 for every
distance with a storage $T_1$ of 1~ms and a 1-$\mu$s cycle; at $d=9$--15 operating at $\nbar=16$ instead costs a factor
1.36 ($\lambda=3$) or 1.6 ($\lambda=1$). A flat CX part gives a plateau instead (within 10\% of its minimum from
$\nbar\approx3$ to 16 at $d\ge5$), whose drifting argmin, from about 7 at $d=5$ to about 13--15 at $d=15$, carries no meaning. Hann
\textit{et al.} note that operating beyond the photon number at which the CX floor takes over ``is thus not
beneficial''~\cite{Hann2025Hybrid}; the growing floor penalizes it. The effect is modest where the code performs best: at
the optimal distance ($d=3$ for $T_1=1$~ms and a 1-$\mu$s cycle) the optimal photon numbers of the three models agree
within half a photon (App.~\ref{app:projection}).

\section{What the published Ocelot data constrain}
\label{sec:ocelot}

\textit{Model.}---We treat each of the eight storage--ancilla interactions of Ocelot~\cite{Putterman2025Hardwareefficient}
as a cell of a storage resonator, a flux-tunable transmon coupler and a fixed-frequency transmon ancilla, in the laboratory
frame without rotating-wave approximation, with circuit parameters inferred from static data only---frequencies,
dispersive shifts and Kerr coefficients---never from error rates (App.~\ref{app:model}, whose Table~\ref{tab:data} lists
every dataset used and the outcome of each comparison). The ancilla's charge couples to a smooth bath referenced to the
measured idle lifetimes and to discrete defects (App.~\ref{app:tls}), with the universal Lindblad
equation~\cite{Nathan2020Universal} in the dressed basis; the gates follow the dressed branches adiabatically:
Landau--Zener passages add $\sim10^{-8}$ per gate, and spectator modes matter only at an accidental resonance within a few
kHz (App.~\ref{app:spectra}).

\begin{figure}
\includegraphics[width=\columnwidth]{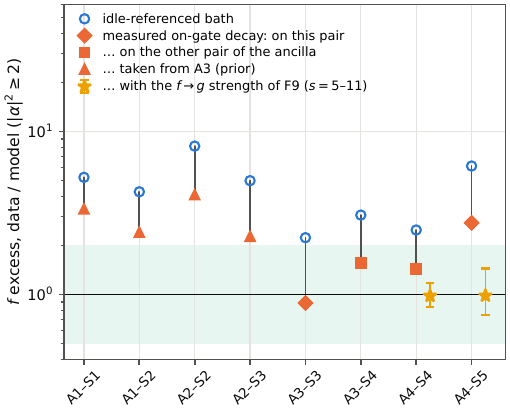}
\caption{CX$^2$ $f$ excess of the eight Ocelot pairs~\cite{Putterman2025Hardwareefficient}, data/model (median over
$|\alpha|^2\ge2$): idle-referenced bath (open circles); with the on-gate decay measured on the pair, on the other pair of
the ancilla, or the coupler loss of A3 as a prior (filled); and, for A4, with the smooth $f\to g$ channel at the strength
favoured by F9, $s=8$ (stars; bars $s=5$--11). Band: within a factor of 2.}
\label{fig:cxratios}
\end{figure}

\textit{Gate errors.}---Ref.~\cite{Putterman2025Hardwareefficient} reports the bit-flip times of repeated CX$^2$ (a full turn
of the $f$ branch every 3~$\mu$s) for all eight pairs at $|\alpha|^2=1$--4, with the ancilla in $g$, $g+e$, $f$ and $g+f$;
the excess with the ancilla in $f$ over $g$ isolates the fault-driven part, once the deterministic Kerr shear of the model
is separated [Fig.~\ref{fig:cxratios}; App.~\ref{app:cx}]. At $|\alpha|^2=2$--4 the data exceed the idle-referenced model
by a factor 4.6 in the median over the pairs (2.2--8.1 between pairs; 4.3 over all points); an undressed model is 8.9
times below. The model singles out the $\chi$-mismatched A4--S4, whose deterministic conversion puts it about four times
above the other pairs in the data, but it does not reproduce the order of the other seven. The level needs faster decay
at the operating point, $\lambda\approx2.7$ in the median over the pairs, close to the $\approx3$ relative to the idle $T_1$
that the experiment's own simulations need (2.5 relative to the $T_1$ measured without
pulsing)~\cite{Putterman2025Hardwareefficient}; published coherence measurements of ancillas A3 and A4 at the operating
point supply part of it and bring the median of the pairs from 4.6 to 2.3. The excess at
$|\alpha|^2\le1.5$, the floor of the $g$ protocol and the $g+e$ protocol of A3--S4 are not explained.

\textit{The share of the opened channel.}---In the stabilizer measurements at $|\alpha|^2=2$--4 the direct $f\to g$ makes
11--66\% of the model's logical-relevant error at $\lambda=1$ and 6--44\% at $\lambda=3$. Against the measured gate errors
it thus contributes a few per cent to 15\% at its idle-referenced strength, and up to about 40\% if it shares the threefold
on-gate enhancement: in the device as built it is a sub-leading error. The published $f$ excess does not grow with
$|\alpha|^2$ at $|\alpha|^2=2$--4, where $c_1\nbar$ is still below $c_0$, so these data do not measure $c_1$.

\textit{The anomalous direct $f\to g$ decay.}---After an 800-ns pulse to the operating point with $\nbar$ photons in
storage S4, ancilla A4 prepared in $f$ decays directly to $g$, with a peak near $|\alpha|^2\approx10$ and the $e$ population
at its baseline (App.~F9 of Ref.~\cite{Putterman2025Hardwareefficient}; ``F9''). In a charge bath this needs the opened
element: its smooth part at the idle-referenced strength gives only $\approx2\%$ of the peak ($\approx5\%$ with the
erasure-based $T_1^{fe}=12.5~\mu$s), and a defect in the 26-MHz band
through which the photons sweep the released quantum makes the peak (App.~\ref{app:tls}). One-defect fits with the smooth
channel free favour $s=5$--11 times its idle-referenced strength, both for a charge-coupled defect and for a junction defect
that modulates $E_J$, which the data do not separate; the on-gate $T_1(f)$ of A4 accounts for a factor $\approx1.6$ of $s$,
the rest would be $\sigma\approx3$--7 near 5.00~GHz. At $s=8$, with nothing adjusted to the gate data, the opened family
carries 27--53\% and 68--76\% of the modelled $f$ excess of A4--S4 and A4--S5, and the model matches both (data/model 0.98;
stars in Fig.~\ref{fig:cxratios}): consistency, not an independent determination.

\textit{Logical memory.}---The public raw data of the repetition-code run~\cite{Putterman2024Data} reproduce the published
logical bit flips~\cite{Putterman2025Hardwareefficient}; the measured single-gate and idle rates account for them within
10\% in the distance-5 code (17\% in the distance-3 sections), and our model with $\lambda=3$, within 0.95--1.4 of the
data, carries the gate-level factor to the logical level. The per-round erasure
flags bound the on-gate $f\to e$ enhancement by $\lambda_f\le2.6$--3.6 and do not depend on $|\alpha|^2$, as
$\chi$-matching predicts (App.~\ref{app:logical}).

\section{Discussion}
\label{sec:discussion}

\textit{What the theory establishes.}---Total parity decides which ancillas the photons of a bosonic code expose to charge
noise: equal-parity ancillas acquire a photon-stimulated Raman decay at a rate relative to $f\to e$ given by
Eq.~\eqref{eq:law} and the spectral factor $\sigma$, qubit ancillas only the dressed dephasing of $\Pi$-even noise. For a
$(g,f)$ transmon $\chi$-matched by its detuning the cost, $\approx\Gamma_{fe}/\alpha$ per gate and photon, is lowered by
neither a weaker coupling nor a faster gate, only by a longer $T_1^{fe}$, a larger $\alpha$, a deliberate mismatch,
matching by a drive or a qubit ancilla. Unheralded and converting with probability $\approx\frac12$, the opened decay is the
charge-noise error that $\chi$-matching leaves at first order; it leads the silent faults once $(\nbar+1)\bar\rho$ exceeds
$\Gamma_{eg}T_{ab}$, and in a repetition code it makes the floor of the gates grow with $\nbar$. These statements rest on
the parity structure, on perturbation theory checked against exact ladders and on a model of the syndrome cycle; the
absolute size of the floor in a device also depends on $\sigma$.

\textit{Predictions.}---The theory differs from a photon-independent floor in three predictions. (i) The CX contribution to
the bit flips per cycle grows 1.2--1.7-fold from $\nbar=4$ to 10 in the Ocelot model, where a constant floor stays flat;
this growth measures $c_1$. (ii) Without a defect in its swept band, an $f$-prepared ancilla held at the operating point
gains a $g$ population that grows smoothly with the photon number: after 800~ns, $\approx3\times10^{-3}$ at $\nbar=20$ for
A4--S4 at the idle-referenced strength ($6\times10^{-3}$ with $T_1^{fe}=12.5~\mu$s), and $s$ times more at the strength
that F9 favours. (iii) A defect anomaly on the
opened channel appears with only one storage of an ancilla, whereas a defect that modulates $E_J$ would appear with both
(App.~\ref{app:tls}).

\textit{Limitations.}---Calibrated without error-rate data, the model needs an on-gate decay about three times the idle one
to reach the level of the gate and logical bit flips, half of which independent coherence measurements supply. Not
explained are the $g$-protocol floor, the excess at $|\alpha|^2\le1.5$, the $g+e$ protocol of A3--S4 and most of the
growth of the measurement error with $|\alpha|^2$ (App.~\ref{app:logical}), and $c_1$ is not measured. The coupler parameters are priors, so
band positions are uncertain by tens of MHz and statements about defects are statistical. The smooth bath is flat over the
$\sim0.5$-GHz window, with channels acting independently; a structured bath turns $c_1$ into a distribution: in the
defect ensembles the 97.5\% quantile of the $f\to g$ probability per CX is 3--20 times the median at the literature
density and 8--23 times at three times its surface density. The gates are adiabatic, the data come from several
cooldowns between which defects reconfigure, the coupler loss is transferred between couplers, and the circuit
parameters of GKP device A are inferred from dispersive data. None of these limits affects the parity structure or the law in measurable parameters; they
set the uncertainty of the absolute rates.

\section*{Data availability}
Three reference noise models of repetition cat codes as Stim circuits~\cite{Gidney2021Stim}---the Ocelot-type hybrid
cat--transmon model built from the fault tables of Sec.~\ref{sec:conversion}, the dissipative-cat model of Le~R\'egent et
al.~\cite{LeRegent2023Highperformance} with a refresh step between the two
CNOTs~\cite{Gouzien2023Performance}, and the Kerr-cat tables of Darmawan et
al.~\cite{Darmawan2021Practical} made parametric in $\nbar$---with 72 syndrome datasets generated from them, and the
package \texttt{bosonoise}, with one command for every figure and data product of this paper, are available
(code under Apache-2.0, data products under CC~BY~4.0 except data digitized from published figures) at
\url{https://github.com/gulkanov/bosonoise}~\cite{Gulkanov2026Bosonoise}; every parameter of the noise models carries its
source. The raw data of the Ocelot repetition-code experiment used in Sec.~\ref{sec:ocelot} and App.~\ref{app:logical}
are publicly available~\cite{Putterman2024Data}.

\begin{acknowledgments}
The author acknowledges support from the Ken Kennedy Institute Computational Science and Engineering Recruiting
Fellowship, funded by the Energy HPC Conference.
\end{acknowledgments}

\appendix

\section{The Ocelot cell: model and calibration}
\label{app:model}

\begin{table*}
\caption{Published data of the Ocelot device~\cite{Putterman2025Hardwareefficient}, and of its companion
device~\cite{Putterman2025Preserving}, used in this work, with the outcome of each comparison. The cell model is
calibrated on static data only, and error rates validate it; the measured idle bit-flip rates and storage lifetimes enter
the logical-level model as inputs, and the on-gate factor $\lambda=3$ is chosen to match the gate data. Ratios are
data/model.}
\label{tab:data}
\centering\small 
\begin{tabular}{lL{4.1cm}L{2.9cm}L{5.8cm}l}
\toprule
source & observable & role & outcome & where\\
\midrule
Fig.~7 & circuit-model curves of A2--S3 against coupler flux & calibration (fit A) & largest pull 0.33, rms 0.10 (0.94
and 0.21 over all 52 targets) & \ref{app:model}\\
Fig.~9 & idle frequencies and $T_1$ of ancillas and storages & calibration; idle-referenced bath & input &
\ref{app:model}\\
Fig.~12 & CX lengths; $\chi_{ge}$, $\chi_{gf}$ at the operating point & calibration; gate pulses & published model and
measured shifts differ by 7--9\%; fit B stays 3--4\% short & \ref{app:model}, \ref{app:cx}\\
Fig.~16 & CX$^2$ bit-flip times, 8 pairs, 4 protocols, $\nbar=1$--4 & validation; static protocol: idle part (input) & $f$
excess, median of pairs: 4.6 (idle-referenced), 2.3 (measured on-gate decay); A4 pairs 0.98 with the $f\to g$ strength of
F9 ($s=8$); A4--S4 singled out, order of the other pairs not reproduced & \ref{sec:consequences}, \ref{sec:ocelot},
\ref{app:cx}\\
Fig.~18c & $T_1$ of A3 against coupler flux & on-gate decay; defect density & coupler $T_1=2.37~\mu$s at the operating
point; defect density one to three times the literature prior & \ref{app:cx}, \ref{app:tls}\\
Fig.~37d & pulsed $T_1(e)$, $T_1(f)$ of A4, with and without photons & on-gate decay & photon-dependent $T_1(e)$ from
Stark-swept defects in 4.9--7.0\% of realizations; $T_1(f)$ 3\% shorter at $s=8$, measured $6\pm6\%$ & \ref{app:cx},
\ref{app:tls}, \ref{app:stark}\\
Fig.~18b (App.~F9) & populations after 800~ns, $\nbar=0$--20 (A4--S4) & direct $f\to g$ decay & one defect plus the smooth
channel fit to shot noise, $s=5$--11 & \ref{sec:ocelot}, \ref{app:tls}\\
Figs.~3f, 4b; raw data & logical phase and bit flips, syndromes, erasure flags~\cite{Putterman2024Data} & validation &
bit flips 1.4--5.4 (idle-referenced), 0.95--1.4 ($\lambda=3$), gate part 3.8--7.2 and 0.9--1.4; phase flips: model within
$-5\ldots+12\%$ of the data (data-anchored); erasures bound $\lambda_f\le2.6$--3.6 & \ref{sec:ocelot}, \ref{app:logical}\\
Fig.~30 & storage lifetimes in the repetition code & phase flips; bias ceiling & storage decays 1.1--1.4 times faster
during the gates than at idle & \ref{sec:consequences}, \ref{app:logical}\\
Fig.~7b of Ref.~\cite{Putterman2025Preserving} & cost of pulsed stabilization & validation & reproduced at
$|\alpha|^2\le3$ without a fit & \ref{app:switch}\\
\bottomrule
\end{tabular}
\end{table*}

\subsection{Hamiltonian}

Each cell consists of a storage resonator $a$, a flux-tunable transmon coupler $c$ and a fixed-frequency transmon ancilla
$q$, with the laboratory-frame Hamiltonian, without rotating-wave approximation,
\begin{multline}
H = f_s\, a^\dagger a + \tfrac{K_{\rm x}}{2}\,a^{\dagger2}a^2 + H_c[\Phi] + H_q \\
+ (g_{sc}\hat n_c + g_{sq}\hat n_q)(a+a^\dagger) + g_{cq}\,\hat n_c\hat n_q .
\label{eq:Hcell}
\end{multline}
$H_q$ and $H_c$ are transmons in the charge basis with the full cosine potential; the coupler has an asymmetric SQUID,
$E_J(\Phi)=E_{J,\max}\sqrt{\cos^2\pi\Phi+d^2\sin^2\pi\Phi}$. $K_{\rm x}$, a flux-independent storage Kerr from the buffer and
its asymmetrically threaded SQUID, is the only spectator term and is fitted as a constant. The product basis of the separately
diagonalized transmons is truncated by levels in the calibrations and by a flux-independent bare-energy cutoff along the
flux path; results are for $n_g=0$, and the branches of A2--S3 and the GKP devices were also checked at $n_g=\frac14$ and
$\frac12$.

\subsection{Calibration on static spectroscopy}

\begin{figure}
\includegraphics[width=\columnwidth]{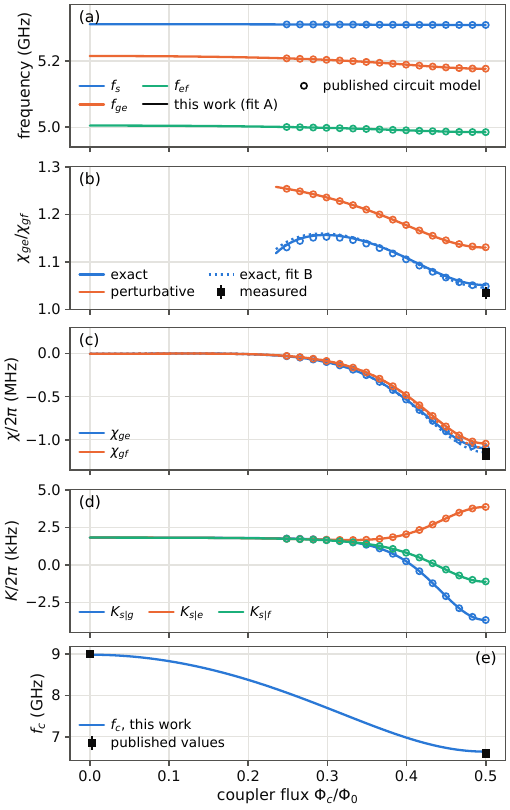}
\caption{Calibration of the A2--S3 cell against coupler flux: model (fit A, lines) and the published circuit-model curves of
Fig.~7 of Ref.~\cite{Putterman2025Hardwareefficient} (circles). (a) Storage and ancilla frequencies. (b)
$\chi_{ge}/\chi_{gf}$, exact and perturbative [Eq.~(D5) of Ref.~\cite{Putterman2025Hardwareefficient}]. (c) Dispersive
shifts. (d) Storage Kerr for the ancilla in $g$, $e$, $f$. (e) Coupler frequency. Dotted, fit B; squares, measured or
published values.}
\label{fig:calibration}
\end{figure}

\begin{table*}
\caption{Calibration of the A2--S3 cell on static spectroscopy. Charge couplings are the coefficients of
Eq.~\eqref{eq:Hcell}; $g_{01}$ are the corresponding couplings of the $0$--$1$ transitions.}
\label{tab:calA2S3}
\centering\small 
\begin{tabular}{lcc}
\toprule
parameter & fit A & fit B \\
\midrule
$f_s$ (bare) & 5.31385 GHz & 5.31377 GHz \\
$E_{J,q}$ / $E_{C,q}$ & 18.868 / 0.19827 GHz & 18.873 / 0.19807 GHz \\
$E_{J,\max,c}$ / $E_{C,c}$ / $d_c$ & 62.11 / 0.1674 GHz / 0.5463 & 61.41 / 0.1700 GHz / 0.5365 \\
$g_{sc}$ / $g_{sq}$ / $g_{cq}$ & 54.38 / 10.61 / 163.24 MHz & 53.42 / 10.11 / 161.55 MHz \\
$g_{01}$ at idle ($sc$ / $sq$ / $cq$) & 99.4 / 13.7 / 384.4 MHz & 97.0 / 13.0 / 377.9 MHz \\
$g_{01}$ at the operating point & 85.2 / 13.7 / 329.3 MHz & 82.7 / 13.0 / 322.2 MHz \\
$K_{\rm x}$ & +1.851 kHz & +1.893 kHz \\
\bottomrule
\end{tabular}
\end{table*}

For the reference cell A2--S3 the 52 targets are the published circuit-model curves of Fig.~7 of
Ref.~\cite{Putterman2025Hardwareefficient} at $\Phi=0.25$--$0.5\,\Phi_0$ (frequencies, $\chi_{ge}$, $\chi_{gf}$ and the storage
Kerr for each ancilla level), the measured idle frequencies and the coupler frequency range (fit A); fit B uses the measured
$\chi$ at the operating point instead (Table~\ref{tab:calA2S3}). Fit A reproduces the curves with pulls of at most 0.33
(rms 0.10) and all 52 targets with a largest pull of 0.94, the measured idle frequency of A2, and an rms of 0.21
(Fig.~\ref{fig:calibration}). Fit B cannot reach the measured $\chi$ (pulls 3.6 and 4.6): the published model gives shifts
7--9\% smaller than measured at the operating point, and no parameter set of the three-mode reduction reaches both, fit B
staying 3--4\% short. The two fits differ by at most 5\% in any parameter, and their spread is our systematic band.

\begin{table*}
\caption{Joint static calibrations per ancilla: cost (half the sum of squared pulls), the largest pulls, idle anharmonicity
(signed; elsewhere $\alpha$ denotes its magnitude) and storage--ancilla detuning at idle and at the operating point.}
\label{tab:calall}
\centering\small 
\begin{tabular}{lcccc}
\toprule
ancilla & cost & largest pulls & $\alpha$ idle (MHz) & detuning idle / on (MHz) \\
\midrule
A1 & 5.02 & $f_{S1}$ $+2.2\sigma$, $f_{S2}$ $-1.9\sigma$ & $-181.3$ & S1 51/92, S2 71/111 \\
A2 & 0.56 & $f_{S2}$ $+0.6\sigma$ & $-201.6$ & S2 73/111, S3 98/135 \\
A3 & 0.74 & $f_{S3}$ $-0.7\sigma$ & $-180.4$ & S3 96/138, S4 152/191 \\
A4 & 4.11 & $f_{S4}$ $-2.1\sigma$, $f_{S5}$ $+1.3\sigma$ & $-214.5$ & S4 27/70, S5 94/135 \\
\bottomrule
\end{tabular}
\end{table*}

The other cells are calibrated jointly per ancilla with the coupler of A2--S3 as a prior (shared $E_J$ and $E_C$ of the
ancilla; per cell $f_s$ and $g_{sc}$), against the measured frequencies and the dispersive shifts at the operating point
(Fig.~12 of Ref.~\cite{Putterman2025Hardwareefficient}). A2 and A3 fit within one standard deviation; A1 and A4 show
$2\sigma$ tensions in their storage frequencies (Table~\ref{tab:calall}), which may come from the unpublished coupler: a
coupler minimum frequency of 6.25 or 6.75 instead of 6.57~GHz changes the fitted A4 parameters and its cost (6.3 and 3.2
against 4.1) but leaves the 26-MHz band of the opened channel, its element (within $\pm20\%$) and the 70-MHz separation of
the A4--S4 and A4--S5 bands unchanged, so statements on absolute band positions are statistical.

\begin{table}
\caption{$\chi$-matching in the Ocelot pairs at the operating point: $x=\Delta/\alpha$ and $\alpha$ from the model,
$r=\chi_{ge}/\chi_{gf}$ from the model and measured (Fig.~12 of Ref.~\cite{Putterman2025Hardwareefficient}), and the cost per
gate and photon of Eq.~\eqref{eq:pergate} with the measured $r$ and $T_1^{fe}=25~\mu$s.}
\label{tab:chimatch}
\centering\small
\begin{tabular}{lccccc}
\toprule
pair & $x$ & $\alpha/2\pi$ & $r$ & $r$ & cost\\
 & & (MHz) & model & meas. & \\
\midrule
A1--S1 & 0.57 & 162.0 & 1.123 & 1.118 & $7.7\times10^{-5}$\\
A1--S2 & 0.68 & 162.8 & 1.056 & 1.065 & $5.7\times10^{-5}$\\
A2--S2 & 0.61 & 182.9 & 1.100 & 1.096 & $6.2\times10^{-5}$\\
A2--S3 & 0.74 & 183.2 & 1.029 & 1.035 & $4.4\times10^{-5}$\\
A3--S3 & 0.85 & 162.4 & 0.979 & 0.990 & $3.9\times10^{-5}$\\
A3--S4 & 1.18 & 162.2 & 0.869 & 0.865 & $2.1\times10^{-5}$\\
A4--S4 & 0.37 & 192.1 & 1.286 & 1.297 & $1.35\times10^{-4}$\\
A4--S5 & 0.70 & 192.4 & 1.038 & 1.030 & $4.5\times10^{-5}$\\
\bottomrule
\end{tabular}
\end{table}

The calibrated cells reproduce the measured $\chi$ ratios of all eight pairs to $\le0.011$ (Table~\ref{tab:chimatch}). The
drive-matched ancillas of Fig.~\ref{fig:channel}(c) are directly coupled transmons with their published $\omega_{ge}$ and
$\alpha$: the static $r=0.75$ of Ref.~\cite{Reinhold2020Errorcorrected} ($x=2.19$, $\alpha/2\pi=137$~MHz, $T_1^{fe}=47~\mu$s)
is reproduced as 0.759, the $r=0.394$ of Ref.~\cite{Rosenblum2018Faulttolerant} ($x=-9.52$, 210~MHz, $T_1^{fe}=23~\mu$s) as
0.395. A larger truncation (storage 8$\to$12, coupler 6$\to$9, ancilla 8$\to$11 levels, charge cutoff 20$\to$30) changes
every calibrated observable by at most 55~Hz.

\subsection{Flux pulse}

For a time-dependent flux we use the irrotational gauge of an asymmetric SQUID~\cite{You2019Circuit},
$H_c(t)=4E_Cn^2-E_{J,l}\cos(\varphi-a\varphi_e)-E_{J,r}\cos(\varphi+b\varphi_e)$ with $\varphi_e=2\pi\Phi/\Phi_0$ and the
capacitance fractions $a$, $b$ taken proportional to the junction energies (not published), which nearly cancels the term
$\propto d\Phi/dt$. The pulse is a flat top with Gaussian edges of the experiment's gate length, its ramp solved so that
the $f$ branch completes the conditional rotation (App.~\ref{app:cx}).

\subsection{Breaking of the parity}

The parity $\mathcal P$ of Sec.~\ref{sec:parity} is broken by offset charge only through the charge dispersion, by small
circuit imperfections (forbidden elements $\sim10^{-2}$ in amplitude are observed in transmons, of unknown
origin~\cite{Sank2016MeasurementInduced}; $10^{-4}$ in rate), and during the flux ramps. A static flux keeps $\mathcal P$:
the SQUID potential is a shifted cosine, $-E\cos(\varphi_c-\theta)$, with $\theta$ growing from 0 at idle to 0.71~rad at
$\Phi_0/2$. A changing $\theta$ adds the parity-odd term $-\dot\theta\,\hat n_c$; on the ramps of the eight pairs $\dot\theta$
stays below $4\times10^{-3}$ of the coupler frequency at the operating point.

\subsection{Baths}

The smooth charge bath is flat over the $\sim0.5$-GHz window of the ancilla transitions and referenced per transition
family to the measured idle $T_1^{ge}$ and $T_1^{fe}$ (Fig.~9 of Ref.~\cite{Putterman2025Hardwareefficient}); defects are
drawn from priors (App.~\ref{app:tls}). Dissipation follows the universal Lindblad equation~\cite{Nathan2020Universal},
$L=\sum_{mn}\sqrt{\Gamma(E_n-E_m)}\,X_{mn}\ket m\langle n|$, with detailed balance at the measured thermal populations and
without Lamb shift. The coupler's own bath at the operating point is inferred from measured coherence (App.~\ref{app:cx}).

\section{Spectra: the opened element and no resonances along the gates}
\label{app:spectra}

\subsection{The opened element in the full cell}

\begin{figure}
\includegraphics[width=\columnwidth]{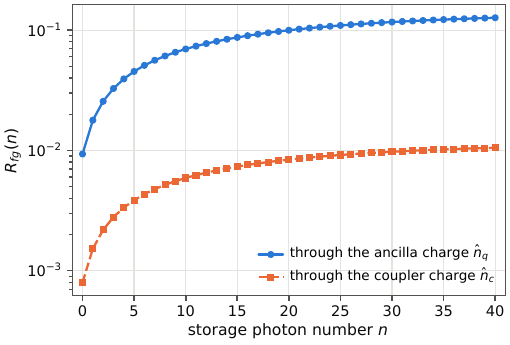}
\caption{Squared opened element of the A4--S4 cell of Ocelot at the CX operating point,
$R_{fg}(n)=\sum_m|\langle\tilde g,m|\hat n|\tilde f,n\rangle|^2$ (dominated by $m=n+1$), through the ancilla charge $\hat n_q$
and the coupler charge $\hat n_c$, relative to the squared bare $f$--$e$ element.}
\label{fig:rfg}
\end{figure}

\begin{table}
\caption{Opened branching per photon at $n=0$, numerically exact in the cell model for the Ocelot pairs and first order
(all levels) for the GKP devices (App.~\ref{app:gkp}), and its ratio to Eq.~\eqref{eq:law}.}
\label{tab:law}
\centering\small
\begin{tabular}{lrrrcc}
\toprule
system & $\alpha$ & $\Delta$ & $\chi_{ge}$ & $\rho$ & ratio\\
 & (MHz) & (MHz) & (kHz) & & to law\\
\midrule
A1--S1 & 162.0 & 92.1 & $-2655$ & $9.02\times10^{-3}$ & 0.98\\
A1--S2 & 162.8 & 111.3 & $-1793$ & $4.58\times10^{-3}$ & 0.96\\
A2--S2 & 182.9 & 110.7 & $-1493$ & $4.01\times10^{-3}$ & 0.95\\
A2--S3 & 183.2 & 134.7 & $-1176$ & $2.35\times10^{-3}$ & 0.94\\
A3--S3 & 162.4 & 137.7 & $-965$ & $1.76\times10^{-3}$ & 0.93\\
A3--S4 & 162.2 & 191.2 & $-1414$ & $1.52\times10^{-3}$ & 0.90\\
A4--S4 & 192.1 & 70.2 & $-1874$ & $9.70\times10^{-3}$ & 0.99\\
A4--S5 & 192.4 & 135.5 & $-1706$ & $3.47\times10^{-3}$ & 0.94\\
GKP A & 222.0 & $-1441.5$ & $-46.6$ & $3.66\times10^{-6}$ & 1.24\\
GKP B & 193.3 & $-1395.9$ & $-34.3$ & $2.39\times10^{-6}$ & 1.21\\
\bottomrule
\end{tabular}
\end{table}

In the full cell the squared opened element grows linearly at small $n$ and more slowly as the states approach the
critical photon number (Fig.~\ref{fig:rfg}): relative to the bare $f$--$e$ element, $R_{fg}=0.0093$, 0.070, 0.100 and 0.127
at $n=0$, 10, 20 and 40 for A4--S4, and 0.0022, 0.021 and 0.036 at $n=0$, 10 and 20 for A2--S3. Divided by the dressed
$f$--$e$ element (0.96 of the bare one), $R_{fg}(0)$ of A4--S4 gives the $\rho(0)=9.70\times10^{-3}$ of Table~\ref{tab:law},
which lists the branching behind Fig.~\ref{fig:channel}(b). The coupler-charge channel is 12 times weaker for A4--S4, 13
times for A4--S5 and 14 times for A2--S3.

\subsection{Branches at the operating point}

\begin{figure}
\includegraphics[width=\columnwidth]{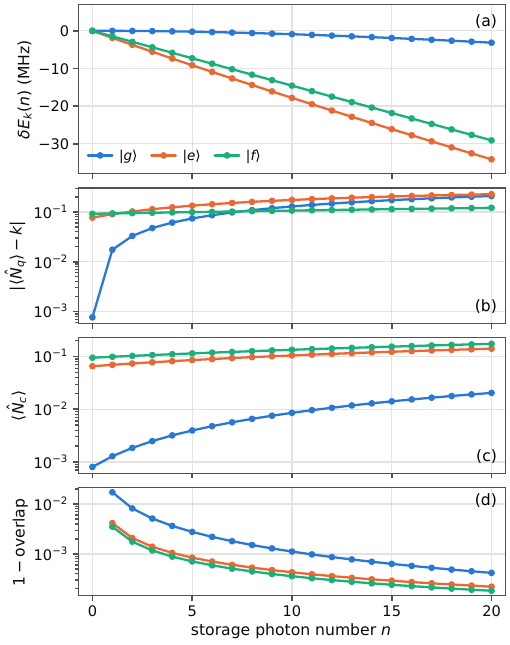}
\caption{Branches of the A4--S4 cell at the operating point for the ancilla starting in $g$, $e$, $f$ ($k=0$, 1, 2). (a)
Energies relative to the linear ladder, $\delta E_k(n)=E_k(n)-E_k(0)-nf_s$, with $f_s$ the $g$-branch storage frequency at
the operating point. (b) Ancilla excitation relative to $k$. (c)
Coupler excitation. (d) One minus the overlap with the normalized $a^\dagger$ applied to the branch state at $n-1$.}
\label{fig:branches}
\end{figure}

We follow the dressed states $\dket{k},n$ (coupler in its ground state) from $n=0$ by maximal overlap with
$a^\dagger\dket{k},n-1$, the static analogue of the branch analysis of measurement-induced
transitions~\cite{Dumas2024MeasurementInduced}. At $\Phi_0/2$ no branch of A2--S3 (fit A at all three offset charges,
fit B at $n_g=0$), A4--S4 or A4--S5 ($n_g=0$) changes character up to $n=20$ (excitation numbers changing by 0.02, or ladder
overlap below 0.98) (Fig.~\ref{fig:branches}); the states are smoothly dressed (ancilla--coupler admixture of the $e$ and
$f$ branches 6--9\% at $n=0$ and 10--23\% at $n=20$, $g$-branch ancilla excitation 0.06--0.13 at $n=10$), so the
dispersive approximation fails gradually, fastest for A4--S4. At the 13 flux points of the gate ladders and up to $n=40$
all eight cells keep branch overlaps $\ge0.976$ and second differences of the branch energies $\lesssim40$~kHz.

\subsection{Crossings along the flux pulse}

\begin{figure*}
\includegraphics[width=\textwidth]{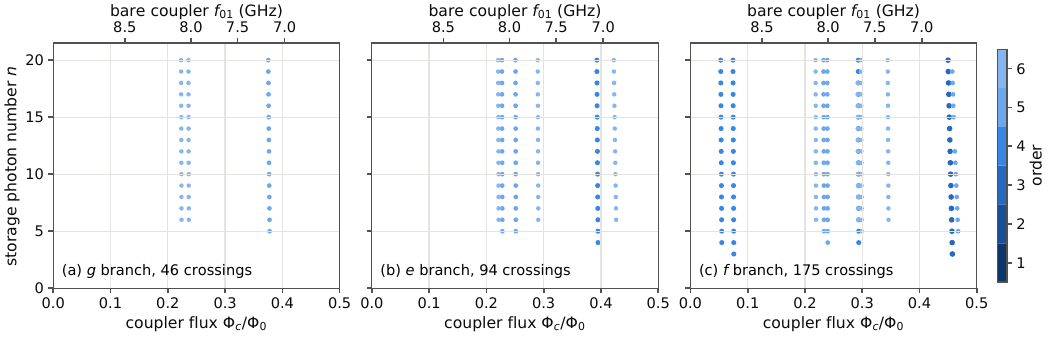}
\caption{Crossings of the $g$, $e$, $f$ branches of the A2--S3 cell with other states of order $\le6$ along the flux
pulse, against the photon number; darker points have lower order. Top axis: bare coupler frequency.}
\label{fig:atlas}
\end{figure*}

Along the flux trajectory A2--S3 has 315 crossings of order $\le6$ with other states (dressed states labelled by their
dominant bare state on a 0.01-$\Phi_0$ grid), in families nearly independent of $n$ (Fig.~\ref{fig:atlas}); the
lowest-order family, $\ket{f,0_c,n}\leftrightarrow\ket{4,1_c,n-3}$ at $\Phi\approx0.45$, has gaps up to 9.9~kHz at $n=20$.
With the single-passage Landau--Zener probability $1-\exp(-\pi\Delta^2/2v)$ on Gaussian edges of $\sigma=16$~ns (longer edges
increase it proportionally), the sum over the 20 families and both passages of a gate is $1.1\times10^{-8}$ at $n=10$ (the
lowest-order family alone $3.5\times10^{-8}$ at $n=20$), five orders below the observed error scale, and still about three
orders below for gaps ten times our resolution-limited bounds.

\subsection{Coupler-excited branches and spectator modes}

An excited coupler shifts the storage by $\chi_{sc}=-0.19$~MHz at idle and $-1.12$~MHz at the operating point (A4--S5), so
an excitation during a gate converts into a bit flip with probability $\approx\frac12$; coupler-excited branches are kept as
a diagnostic. Among the spectator modes (readout resonators, buffers, the idle second coupler) the lowest coincidences
within 60~MHz are of fifth order, all with the second coupler (effective couplings estimated at 0.3--10~kHz). At exact
resonance they would transfer up to $\sim10^{-3}$ per gate, but they need a coincidence within a few kHz while the dressed
shifts move in MHz steps with $n$: not a generic mechanism, although an accidental coincidence cannot be excluded without
the second coupler's frequency.

\section{Gate simulation and the CX\texorpdfstring{$^2$}{2} protocols}
\label{app:cx}

\subsection{Simulation}

A CX turns the cat by $\pi$ conditioned on the ancilla in $f$; the characterization experiments use CX$^2$, a full turn,
every 3~$\mu$s. We follow the dressed branches $\dket{k},n;\Phi(t)$ adiabatically (coupler in its ground state, or excited
as a diagnostic) on ladders at 13 flux points (neighbouring overlaps of the coupler-ground branches $\ge0.977$ up to $n=17$
for every pair, so no resonance is crossed), with phases fixed so that coherent states stay coherent, in the frame of the
$g$-branch storage frequency that the experiment calibrates. The flat-top pulse has the experiment's CX$^2$ length
(published for A1--S1, derived from the CX lengths of Fig.~12 for the others) and a ramp solved for one full turn of the
$f$ branch; with the stated 64-ns ramps of A1--S1 and A4--S4 the model's gate lengths are within 4\% of these, and the
phase corrections, calibrated at $|\alpha|^2=1$ as in the experiment, are $\le0.002$~rad. The universal-Lindblad jump
operators of $\hat n_q$, $\hat n_c$ and the number operators are grouped into families of fixed ancilla transition and
photon change, without cross-family coherences; integration is split-step with exact dressed phases ($dt=0.5$~ns), photon
cutoffs 15 and 38 give the same logical-relevant error to $10^{-3}$ at $|\alpha|^2\le3$ and to 1\% at 4, and after the gate
the cat is re-stabilized (App.~\ref{app:conversion}).

The conditional storage Kerr at the operating point ($g$ branch: $-40$~kHz for A1--S1, $-24$~kHz for A4--S4, $-2.4$ to
$-15$~kHz for A1--S2, A2--S2, A2--S3, A3--S3 and A4--S5; $f$ branch, where evaluated, 3--10 times smaller) shears the cat
during the unstabilized gate: a deterministic, photon-number-dependent contribution [in the $g$ protocol
$(7.2,7.6,3.6,1.3)\times10^{-4}$ at $\nbar=1$--4 for A1--S1] that separates exactly from the fault-driven part compared
with the data. Recalibrating the linear phase at each $\nbar$, as Ref.~\cite{Putterman2025Hardwareefficient} suggests,
moves it by at most 0.06~rad and changes the $g$-protocol error by at most 10\%: the shear is nonlinear in the phase.

\subsection{The \texorpdfstring{$f$}{f} protocol}

\begin{figure*}
\includegraphics[width=\textwidth]{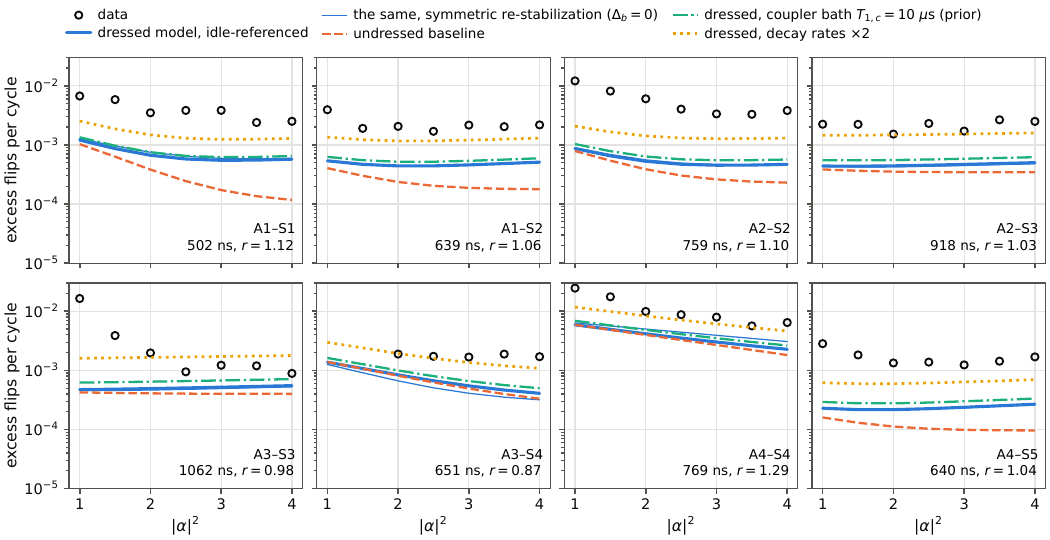}
\caption{Fault-driven excess bit flips per 3-$\mu$s cycle with the ancilla in $f$ (data: Fig.~16 of
Ref.~\cite{Putterman2025Hardwareefficient}) against the dressed model with the idle-referenced bath (thick), the same with
a symmetric re-stabilization map (thin), the undressed baseline (dashed), a coupler bath of $T_1=10~\mu$s (dash-dotted, a
prior) and doubled decay rates (dotted, a diagnostic). Labels: gate length and the model's $r$.}
\label{fig:cxule}
\end{figure*}

\begin{figure*}
\includegraphics[width=\textwidth]{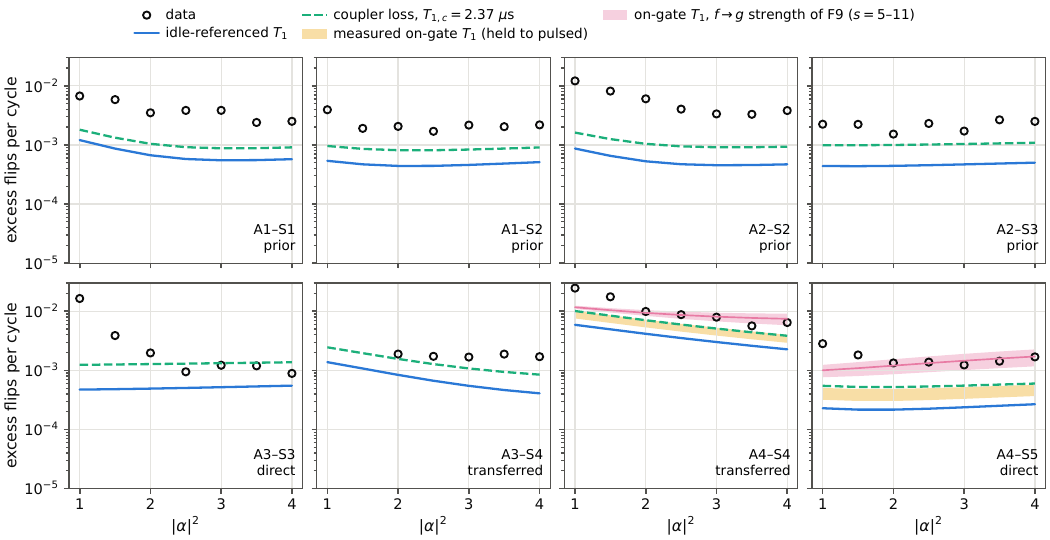}
\caption{Excess bit flips with the ancilla in $f$ (data: Fig.~16 of Ref.~\cite{Putterman2025Hardwareefficient}) against the
model with the idle-referenced $T_1$ (solid), the coupler loss inferred from Fig.~18c (dashed) and, for A4, the on-gate
$T_1$ of Fig.~37d (band: coupler held on to pulsed with $|\alpha|^2=4$) plus the F9 strength of the smooth $f\to g$ channel
(second band, $s=5$--11; line, $s=8$). Labels: whether the measured on-gate decay is that of the pair itself (direct), of
the other pair of its ancilla (transferred: Fig.~18c for A3, Fig.~37d for A4) or the coupler loss of A3 (prior).}
\label{fig:measured}
\end{figure*}

The published bit-flip times are inverse Pauli rates, $p=[1-e^{-2T_c/T_X}]/2\approx T_c/T_X$ per cycle; the excess with
the ancilla in $f$ over $g$ cancels the idle errors (Figs.~\ref{fig:cxule} and \ref{fig:measured}). Beyond the medians of
Sec.~\ref{sec:ocelot}, the ratios span 1.6--11.4 over the 40 points at $|\alpha|^2=2$--4 (2.2--22 for the undressed model),
and the $\lambda$ that matches each point spans 1.4--5.0. The experiment's simulations need a factor $\approx2.5$ relative
to the $T_1$ measured without pulsing~\cite{Putterman2025Hardwareefficient}, $\approx3.1$ relative to the idle $T_1$, where
our model needs 3.1--3.7 for A4--S5. In the data only A4--S4 falls with $|\alpha|^2$ as in the model; A3--S4 ($r=0.87$)
stays at 1.7--1.9$\times10^{-3}$ while the model falls by half, and the matched pairs scatter by up to a factor 2 without a
common trend. The asymmetric re-stabilization map makes A4--S4 17--26\% lower and A3--S4 26--34\% higher than a symmetric
one, consistent with the experiment's simulation for A4--S4. At $|\alpha|^2\le1.5$ the data exceed the model much more
(A3--S3: $1.6\times10^{-2}$ at $|\alpha|^2=1$, 35 times the model), which is not understood; the comparison uses
$|\alpha|^2\ge2$.

Part of the faster on-gate decay is measured. The $T_1$ of A3 against coupler flux (Fig.~18c of
Ref.~\cite{Putterman2025Hardwareefficient}), a median of $38.6~\mu$s for $\Phi\le0.2\,\Phi_0$ and $22.3~\mu$s near $\Phi_0/2$,
requires in our model a coupler $T_1$ of $2.37~\mu$s at the operating point, acting through the 7--12\% coupler admixture of
the ancilla transitions. The pulsed $T_1$ of A4 (Fig.~37d, on A4--S5) is $T_1(e)=34.3$, 30.3, $23.9~\mu$s and $T_1(f)=20.3$,
16.0, $15.1~\mu$s with the coupler held at its operating point, pulsed (one CX$^2$ per 1.6~$\mu$s) and pulsed with
$|\alpha|^2=4$, to which we scale the on-gate rates. The data/model ratio then becomes 0.9 and 2.8 for the pairs measured
(A3--S3, A4--S5), 1.6 and 1.4 for the other pair of the same ancillas, and 2.3--4.1 for A1 and A2 with the coupler loss of
A3 as a prior (Fig.~\ref{fig:cxratios}). Coupler loss alone does not suffice: a single coupler $T_1\ge1.5~\mu$s fails for 21
of the 24 points at $|\alpha|^2=2$, 3, 4, and the faster storage decay during the gates would need 0.1--1.2~$\mu$s. For A4--S5 the remaining excess
comes from shots that end with the ancilla in $g$ (App.~J of Ref.~\cite{Putterman2025Hardwareefficient}), which
App.~\ref{app:tls} compares with the F9 strength.

\subsection{The \texorpdfstring{$g+e$}{g+e} protocol}

The $g+e$ protocol leaves the equal mixture of the $g$ and $e$ branches, so $p(g+e)-p(g)=[p_e-p_g]/2$. For a positive $\chi$
mismatch the deterministic rotation dominates (for A1--S1 at $|\alpha|^2\le3$): A4--S4 (data $353$, $216$,
$228\times10^{-3}$ at $|\alpha|^2=2$, 3.5, 4) and A1--S1 agree within 1.1--1.6 without a fit (0.8--1.3 with the symmetric
map), which validates the conversion of large angle errors. A3--S4 ($r=0.87$) is 1.5--2.8 times low (2.5--3.8 with the
symmetric map), which we do not understand (its $\chi$ ratio, the $e$-branch Kerr shear or state preparation into $e$). For
the matched pairs the protocol measures an $e\to g$ rate during the gate enhanced over idle, by 1.1--1.4 for A2--S3 and
A3--S3, 1.7--2.2 for A4--S5, 2.4--4.3 for A2--S2 (above 5 at $|\alpha|^2=2$) and about 5 or more for A1--S2 at
$|\alpha|^2\ge2$, not growing with the photon number.

\subsection{The \texorpdfstring{$g$}{g} protocol}

With the ancilla in $g$ the fault-driven part at zero temperature is $\le10^{-5}$ (two-photon heating), but the data show a
flat, ancilla-specific floor of $0.23$--$0.84\times10^{-3}$ per cycle at $|\alpha|^2=3$--4 (slope $+0.05\pm0.04\times10^{-3}$
per photon), which as ancilla heating would need 66--93~mK. For A4--S4, storage--ancilla exchange on the up-ramp
(adiabatic perturbation theory) gives $0.30$--$0.62\times10^{-3}$ at $|\alpha|^2=2$--4 with the stated 64-ns ramp
($0.9$--$1.8\times10^{-3}$ with the solved 54-ns ramp), the size and trend of the measured excess over the static
reference, $0.19$--$0.64\times10^{-3}$; exponentially sensitive to the idle detuning and the waveform, it is a candidate,
not a prediction. A2--S2, with the shortest ramp (15~ns), gets $0.03$--$0.15\times10^{-3}$, falling with $|\alpha|^2$; the
other five pairs $\le2.3\times10^{-6}$.

\section{Re-stabilization and fault attribution}
\label{app:conversion}

\subsection{Re-stabilization map}

The two-photon dissipation returns the storage to the code space after each gate. We use the asymptotic projector of
$D[a^2-\alpha^2]$~\cite{Albert2014Symmetries,Albert2016Geometry}, from the left and right null vectors of the Liouvillian,
and count a bit flip as the population of $\ket{1_L}=(\ket{C_+}-\ket{C_-})/\sqrt2$ after the map, starting from $\ket{0_L}$;
the map is exact for the code space: $P(0)=0$, $P(\pi)=1$, $P(\pi/2)=\frac12$, $P(\theta)+P(\pi-\theta)=1$. A detuned buffer
adds a commensurate Kerr-cat term $K(a^{\dagger2}-\alpha^{*2})(a^2-\alpha^2)$, $K/\kappa_2=-\Delta_b/\kappa_b$, which keeps the
steady states but makes the map asymmetric in the sign of a phase error: with the estimated buffer detuning
($\Delta_{b,0}/2\pi\approx-2$~MHz) and the measured linewidths~\cite{Putterman2025Hardwareefficient},
$K/\kappa_2=0.10$--$0.20$, and at
$\nbar=2$ and $K/\kappa_2=0.1$ a phase error of $+0.5$~rad flips with probability 0.011 against 0.019 for $-0.5$~rad. We use
the asymmetric map, the symmetric one as a sensitivity case. Switching the stabilization off during the gates adds no bit
flips of its own (App.~\ref{app:switch}).

\subsection{Fault attribution}

\begin{figure*}
\includegraphics[width=\textwidth]{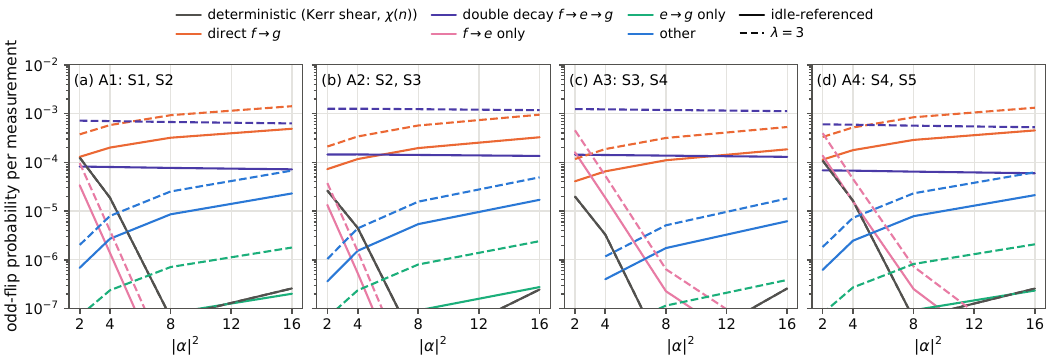}
\caption{Logical-relevant error per stabilizer measurement (an odd number of cat flips) by fault sequence against
$|\alpha|^2$, for the four stabilizers: deterministic (Kerr shear, $\chi(n)$), direct $f\to g$, double decay $f\to e\to
g$, single $f\to e$, single $e\to g$, and other; idle-referenced bath (solid) and $\lambda=3$ (dashed).}
\label{fig:faultconversion}
\end{figure*}

A multi-class first-order Dyson expansion splits the state inside the integrator into components with zero or one jump of
each class ($f\to e$, $e\to g$, direct $f\to g$ with any photon change, heating); unlike a single-class split, which
counts $f\to e\to g$ under both of its jumps, it makes the sequences disjoint, and the components close on the full
evolution to $10^{-12}$. Per stabilizer measurement, ``deterministic $+$ other $+\sum$ sequences'' equals the total in all
48 configurations tested (four stabilizers; idle-referenced, $\lambda=3$ and 50-mK baths; $|\alpha|^2=2$, 4, 8, 16) and in
the 16 of an ancilla three times longer-lived (Fig.~\ref{fig:faultconversion}). The estimate
$(\nbar+1)\bar\rho/(\Gamma_{eg}T_{ab})$ of Sec.~\ref{sec:conversion} uses the branching of Table~\ref{tab:law}, and the
idle $T_1^{ge}$ and the CX lengths of Figs.~9 and 12 of Ref.~\cite{Putterman2025Hardwareefficient}; the tables exceed it by
1.13--1.77 in all 36 configurations (four
stabilizers, $\lambda=1/3$, 1, 3, $|\alpha|^2=4$, 8, 16), slightly less at larger $|\alpha|^2$ as the branching saturates.

\subsection{One stabilizer measurement}

Propagating both ancilla branches through both gates, with the first gate resolved by the ancilla level, and
re-stabilizing each cat with its own map gives the joint probabilities of the two cats' errors, the herald and the
syndrome. In the $f$ branch an $X$ error on the first cat comes mostly with one on the second (82--100\% of the cases at
$|\alpha|^2=2$--4 for A1--A3, 43--90\% for A4), because an early decay to $g$ leaves both cats unrotated (a correlated
$X_aX_b$, trivial on the logical $Z$); the logically relevant single-cat errors thus sit mainly on the second cat. Heralded
events rarely carry cat errors, $P(X|{\rm herald})\lesssim1\%$ at $|\alpha|^2\ge2$; with $\lambda=3$ the
logical-relevant error per measurement at $|\alpha|^2=4$ is $1.2$--$1.6\times10^{-3}$.

\section{Two-level defects and the anomalous direct \texorpdfstring{$f\to g$}{f to g} decay}
\label{app:tls}

Two-level defects (TLS) make the bath structured; because the dispersive shift sweeps every ancilla line with the photon
number, a defect near a line becomes a photon-number-dependent decay channel.

\subsection{One defect in a dressed channel}

\begin{figure*}
\includegraphics[width=\textwidth]{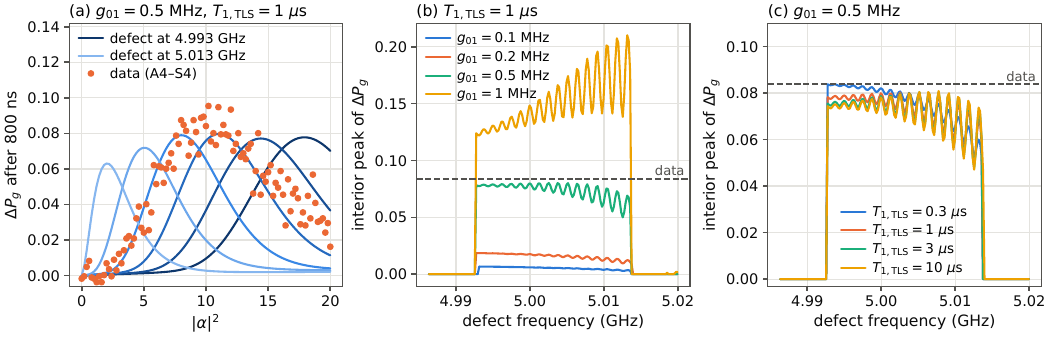}
\caption{One defect in the band swept by the opened channel of A4--S4. (a) $\Delta P_g$ after 800~ns at the operating
point for a defect ($g_{01}=0.5$~MHz, $T_{1,\rm TLS}=1~\mu$s) at six positions across the band, with the data of Fig.~18b
of Ref.~\cite{Putterman2025Hardwareefficient}. (b), (c) Interior peak height against the defect frequency for several
couplings and lifetimes; dashed, the measured peak (0.084).}
\label{fig:tlsfamily}
\end{figure*}

A defect at $f_T$ couples to the ancilla charge with $g_{01}$ (bare $0$--$1$ element) and to a dressed channel of element $M$
as $V=g_{01}|M|/|n_{01}|$. The transfer of each channel $\ket{\tilde i,n}\to\ket{\tilde j,m}$ is computed exactly in the
two-state model $\{\ket{\tilde i,n;0_T},\ket{\tilde j,m;1_T}\}$ with the second state decaying at the defect's rate (the
golden rule at weak coupling, a damped exchange at strong coupling); channels act independently. The released frequency of
the opened channel of A4--S4 moves from $5.016$ to $4.990$~GHz for $n=0$--20, so a defect anywhere in that band is met at
some photon number. For the F9 protocol, a defect with $g_{01}=0.5$~MHz and $T_1=1~\mu$s gives a Poisson-shaped interior
peak of $0.05$--$0.08$ in $\Delta P_g$ ($0.063$--$0.080$ for the six positions of panel (a)) at a position that follows the
defect frequency; the height depends mainly on
$g_{01}$ and little on the defect's lifetime (Fig.~\ref{fig:tlsfamily}). The measured anomaly, 0.084 at $|\alpha|^2\approx10$
(seven-point mean; largest point 0.095) with $P_e$ at its baseline, needs about 0.5~MHz for $T_1=1~\mu$s.

\subsection{Which defect? One-defect fits of F9}

\begin{figure*}
\includegraphics[width=\textwidth]{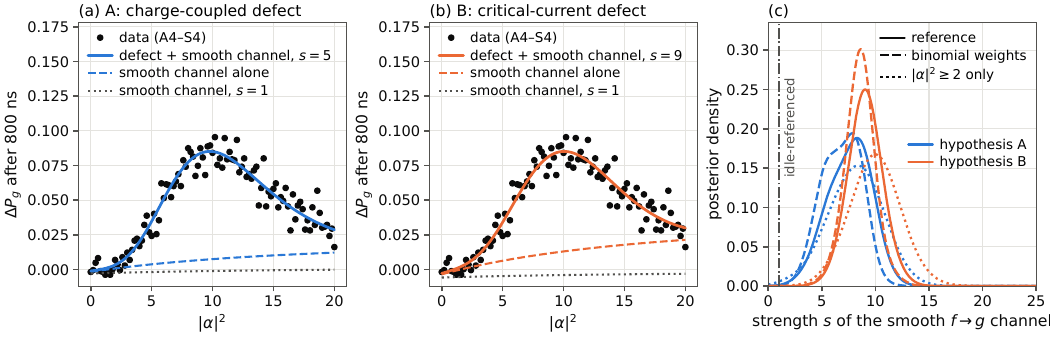}
\caption{One-defect fits of F9 (Fig.~18b of Ref.~\cite{Putterman2025Hardwareefficient}) with the smooth channel free. (a)
Opened channel and a charge-coupled defect (hypothesis A). (b) Critical-current defect (hypothesis B). Lines: maximum a
posteriori (defect plus smooth channel), the smooth channel alone at that strength, and at the idle-referenced strength
$s=1$. (c) Posterior of $s$ under both hypotheses, for the reference fit, binomial noise weights and the points at
$|\alpha|^2\ge2$ only.}
\label{fig:bayes}
\end{figure*}

\begin{table*}
\caption{Defect statistics. Left: the observed three dips of the Fig.~18c scan under each prior (4000 realizations).
Right: fraction of realizations with an F9-sized anomaly (interior peak $\ge0.05$) per pair and cooldown (2000
realizations).}
\label{tab:density}
\centering\small 
\begin{tabular}{lccc|lcc}
\toprule
prior & mean dips & $P(N=3)$ & $P(N\ge3)$ & prior & A4--S4 & A4--S5\\
\midrule
broad, 10/GHz & 1.16 & 0.029 & 0.13 & broad, 20/GHz & 15\% & 12\%\\
broad, 20/GHz & 2.48 & 0.051 & 0.27 & broad, 40/GHz & 26\% & 23\%\\
broad, 40/GHz & 4.82 & 0.073 & 0.48 & broad, 80/GHz & 41\% & 37\%\\
broad, 80/GHz & 9.70 & 0.066 & 0.76 & literature ($\times1$) & 0.4\% & 0.6\%\\
literature ($\times1$) & 2.48 & 0.029 & 0.16 & literature ($\times3$ surface, 0.05~$\mu$m$^2$) & 0.8\% & 0.6\%\\
literature ($\times3$) & 7.03 & 0.064 & 0.47 & literature ($\times3$ surface, 0.15~$\mu$m$^2$) & 1.7\% & 2.2\%\\
\bottomrule
\end{tabular}
\end{table*}

\begin{table*}
\caption{Best logical error per cycle over $d$ and $\nbar$ (in brackets: $d$, $\nbar^*$) for four scenarios of storage
$T_1$ and cycle time.}
\label{tab:projection}
\centering\small 
\begin{tabular}{lccc}
\toprule
scenario & idle-referenced & $\lambda=3$ & phenomenology \\
\midrule
as built (57~$\mu$s, 2.8~$\mu$s) & $8.7\times10^{-3}$ (15, 3.0) & $1.41\times10^{-2}$ (7, 2.0) & $1.49\times10^{-2}$ (5, 1.75)\\
$T_1=300~\mu$s & $1.5\times10^{-3}$ (9, 5.25) & $4.2\times10^{-3}$ (5, 3.75) & $5.0\times10^{-3}$ (3, 2.5)\\
$T_1=1$~ms & $7.7\times10^{-4}$ (5, 5.25) & $2.2\times10^{-3}$ (3, 3.75) & $2.5\times10^{-3}$ (3, 4.0)\\
$T_1=1$~ms, $T_c=1~\mu$s & $4.4\times10^{-4}$ (3, 4.25) & $1.6\times10^{-3}$ (3, 4.0) & $1.9\times10^{-3}$ (3, 4.5)\\
\bottomrule
\end{tabular}
\end{table*}

\begin{figure*}
\includegraphics[width=\textwidth]{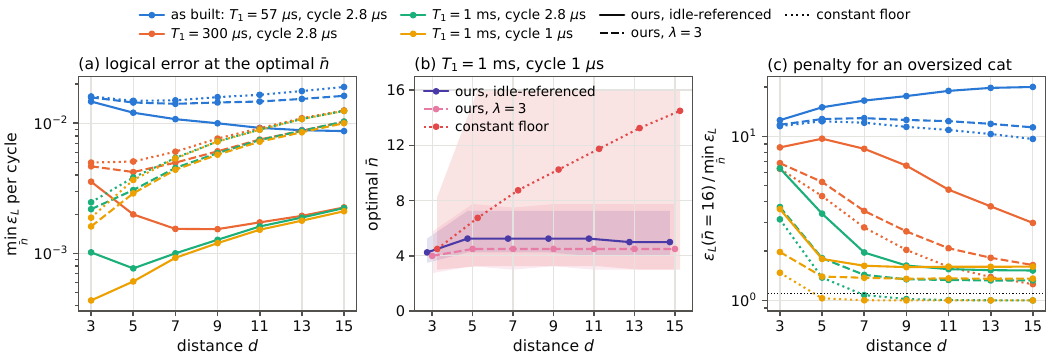}
\caption{Projection of the repetition cat code with transmon ancillas. (a) Minimum over $\nbar$ of the logical error per
cycle against $d$, for four scenarios (colours) and three bit-flip models (line styles). (b) Optimal $\nbar$ for
$T_1=1$~ms and a 1-$\mu$s cycle, with the band in which $\varepsilon_L$ lies within 10\% of its minimum. (c) Penalty for
operating at $\nbar=16$ instead of the optimum.}
\label{fig:projection}
\end{figure*}

We fit F9 with one defect under two hypotheses with identical priors: (A), the mechanism of this paper, a charge-coupled
defect near 5.00~GHz that drives the opened channel through $M_{fg}(n)$, whose element grows from $0.13$ to $0.44$ of
$|n_{01}|$ between $n=0$ and 20; and (B), a parity-even alternative, a junction defect that modulates $E_J$, couples through
$\cos\varphi_q$ (element 0.094--0.091, bare 0.0997) and drives $\ket{\tilde f,n}\to\ket{\tilde g,n}$ at
$\omega_{gf}/2\pi\approx10.35$~GHz. Both lines move by $-1.3$~MHz per photon. The smooth opened channel, which sets $c_1$, is
scaled by a free factor $s$ ($s=1$ idle-referenced), because the on-gate decay is faster than at idle and $\sigma$ need not
be 1. Priors: defect frequency uniform over the band $\pm6$~MHz; coupling at $n=10$ (0.01--1~MHz),
$T_{1,\rm TLS}$ (0.1--10~$\mu$s) and diffusion width (0.03--3~MHz) log-uniform; $s$ uniform in 0--30; offset and noise level
marginalized analytically; evidences by grid quadrature (a coarse grid reproduces a fine one to 0.2 in $\ln Z$ and to 0.01
in $\ln B({\rm A/B})$). Both hypotheses describe F9 to its shot noise (residual rms 0.008, the binomial scatter at
$P_g\approx0.1$; uncorrelated residuals; Fig.~\ref{fig:bayes}) and favour a strong smooth channel, $s=7.7$ ($5.4$--$9.6$,
68\%) under A and $9.0$ ($7.4$--$10.6$) under B, by Bayes factors of $e^{3.2}$ and $e^{8.0}$ over $s=1$, also with binomial
weights and without the baseline region (across these variants and priors on $s$, $e^{0.6}$--$e^{3.9}$ under A and
$e^{3.6}$--$e^{9.1}$ under B): the smooth channel makes the slow rise and the tail, the defect the peak ($g_{01}=0.5$--1.5~MHz under A;
$\delta E_J\sin\theta=2.1$--5.4~MHz under B). The data do not separate the two, $\ln B({\rm A/B})=0.0$--$0.7$; only with $s$
fixed at 1 do they favour A ($\ln B=5.2$, 4.7--7.7 across priors), because A's element grows with $n$ and mimics part of
the tail.

\looseness=1 
A second, broad defect at $s=1$ instead of the stronger smooth channel reaches a slightly higher likelihood but a lower
evidence ($\ln B=1.1$--2.0 across hypotheses, grids and priors), with its most probable positions at $n\approx14$--19. Only a
junction defect can modulate $E_J$; at 0.5--1 junction defects per GHz the 21-MHz window of B holds one with probability
$\le1$--2\% per pair, and its critical-current coupling would have to reach a few MHz ($\delta E_J/E_J\sim10^{-4}$), which has
not been observed; its longitudinal part, $\lesssim0.9$~MHz here, lies below the $\approx3$-MHz detection threshold of
Ref.~\cite{Lisenfeld2026Readout}, which found none.

\looseness=2 
The hypotheses differ on the other storage of A4, the basis of prediction (iii) of Sec.~\ref{sec:discussion}: B's line
starts at the same $\omega_{gf}$ with S5 ($10.3644$ against $10.3645$~GHz at $n=0$, equal within 0.15~MHz for all three
coupler priors) and moves by $-1.60$~MHz per photon, so B's defect would appear with S5 near $|\alpha|^2\approx8$--9 if the
two couplers shift the ancilla's on-point frequency alike within about $-14$ to $+18$~MHz, whereas A's S5 band lies 40--97~MHz below its S4 band (our calibration
of A4 leaves $f_{S4}$ 11~MHz low, a measure of the plausible difference). The strength that F9 indicates is not a
measurement for the code: F9 comes from another cooldown and one pair, and its populations are single-shot classifications
read with the coupler back at idle, where the dispersive shifts ideally vanish, so a photon-number-dependent readout error
is unlikely but not quantified.

\subsection{The F9 strength in the gate data and in the floor}

\looseness=1 
With the opened family of the A4 pairs at the strength $s$ that F9 favours, the other channels at the measured on-gate
values and nothing adjusted to the gate data, data/model is 1.18, 0.98, 0.84 for A4--S4 and 1.44, 0.98, 0.75 for A4--S5 at
$s=5$, 8, 11 (Figs.~\ref{fig:cxratios} and \ref{fig:measured}), and 1.23, 1.02, 0.87 and 1.53, 1.02, 0.77 with the on-gate
$T_1(e)$ and $T_1(f)$ measured without photons, which avoids counting part of this decay twice. The two measurements, from different
cooldowns, are thus consistent with one enhanced strength without determining it: the gate data alone do not separate $s$
from the on-gate decay ($s=11$ with the idle-referenced bath gives 1.05 and 0.80), $s$ refers to $T_1^{fe}=25~\mu$s and
would halve for the erasure-based $12.5~\mu$s, and the A4--S5 band lies 40--97~MHz below the one F9 probes. At $s=8$ the
pulsed $T_1(f)$ at $|\alpha|^2=4$ shortens by $\approx3\%$, against the measured $6\pm6\%$; and the strength is not
universal, A3--S3 being at 0.9 with its measured coupler loss alone. At $s=8$ the cats that A4 serves have
$c_1=2.5\times10^{-4}$ and a bias ceiling of 193 (S5, which only A4 serves) and $1.1\times10^{-4}$ and 424 (S4); 142--305 and
332--587 for $s=5$--11.

\subsection{Ensembles}

\looseness=1 
The literature prior combines surface defects, whose cumulative density per unit frequency falls as $g^{-2}$, normalized to
a defect-limited $T_1$ of $100~\mu$s~\cite{Weeden2026Statistics}, with junction defects with $dN/dg\propto g^{-1}$ up to
30~MHz for a junction area of $0.05~\mu$m$^2$ ($0.15~\mu$m$^2$ as a
variant)~\cite{Martinis2005Decoherence,Muller2019Towards}; realizations whose idle rates exceed the measured ones by more
than a factor 1.5 are rejected, and the smooth background is reduced by the mean rate of the explicit surface defects. A broad prior (density per GHz, couplings log-uniform in 0.02--2~MHz, lifetimes in 0.3--10~$\mu$s)
serves as a sensitivity case; spectral diffusion of 1--5~MHz changes none of the statistics.

\looseness=1 
The flux scan of Fig.~18c of Ref.~\cite{Putterman2025Hardwareefficient}, which moves the $g$--$e$ frequency of A3 over
43~MHz ($5.210\to5.167$~GHz), is a coarse defect spectroscopy: three of its 30 reliable points fall below half of the
smooth trend, which the literature prior reproduces at one to three times its density (mean count 2.5 at its default,
probability of exactly three dips highest at three times) and the broad prior at 20--80 defects per GHz (probability
within a factor 1.5 of its maximum) (Table~\ref{tab:density}). At such densities an F9-sized anomaly occurs per pair and cooldown with probability
0.4--2\% under the literature coupling law or 12--41\% under the $1/g$ law, which the scan does not exclude (had all eight
pairs been scanned in two cooldowns, at least one anomaly would be expected with probability 6--29\% or 87--100\%), and the
$\ge2.8$-fold $f\to g$ enhancement that the A4--S5 residual needs occurs in $\approx10$--17\% of the literature realizations
at three times the surface density. Defects do not raise the typical on-gate decay: the ratio of on-gate to idle decay has
a median of 0.6--0.9 (below 1 through the dressing of the charge elements and the subtraction of the mean defect rate from
the smooth background) and exceeds 1.5 in 7--8\% of the realizations at the literature density, so defects explain
outliers, not the systematic $\lambda\approx2$--3. A defect near the Stark-swept $e\to
g$ line also accounts for the photon-dependent $T_1(e)$ of A4 in Fig.~37d of Ref.~\cite{Putterman2025Hardwareefficient}
(App.~\ref{app:stark}).

\section{Projection to larger codes}
\label{app:projection}

We project to $d=3$--15 with our stabilizer tables up to $|\alpha|^2=16$, the data-derived idle part and a Stim table of the
phase-flip memory ($\varepsilon_{\rm phase}(d,p_Z)$; threshold $p_Z\approx0.2$), for four hardware scenarios
(Table~\ref{tab:projection}, Fig.~\ref{fig:projection}), against the phenomenology with its $|\alpha|^2$-independent CX term
and the same idle part. At a common distance (the optimum for $\lambda=3$) the optimal photon numbers of the three models
agree within half a photon, and the phenomenology's minimum logical error lies 2--27\% above ours at $\lambda=3$ at equal
distance (6--18\% at the best distance of each model). The rise of the minimum logical error with $d$ in
Fig.~\ref{fig:projection}(a) comes from the transmon-ancilla CX floor whether or not it grows with $\nbar$ (App.~\ref{app:floor}).

\section{GKP devices: Floquet--Markov treatment}
\label{app:gkp}

GKP qubits stabilized with echoed conditional displacements (ECD) use a qubit ancilla $(g,e)$, far detuned from the
cavity, and drive the cavity to hundreds of photons during their large
displacements~\cite{Eickbusch2022Fast,Sivak2023Realtime,Brock2025Quantum}, beyond the dispersive regime of
Eq.~\eqref{eq:law}. An
$f$ level would reach only $\rho\approx3\times10^{-6}$ per photon there [Fig.~\ref{fig:channel}(d)]; their protection is
therefore the qubit ancilla, not only the weak dispersive coupling. A Floquet--Markov treatment of two published devices
confirms the parity rule beyond the dispersive regime: the one-photon sidebands of $e\to g$ vanish, those of $f\to g$ are
the opened channel, and below the ionization thresholds the decay of $\ket e$ and the heating of $\ket g$ do not depend on
the photon number.

\textit{Devices.}---Device A~\cite{Sivak2023Realtime,Brock2025Quantum} (parameters of Ref.~\cite{Sivak2023Realtime}) has a
cavity at 4.479~GHz and a transmon at 5.921~GHz ($-222$~MHz, $T_1=280~\mu$s, thermal population 4.3\%,
$|\chi|=46.5$~kHz), with large ECD displacements of 121--256 photons (DAC limit 676); its circuit parameters, derived from
the dispersive data, which they then reproduce within 6--10\%, are
$E_J=23.05$~GHz, $E_C=203.9$~MHz and $g=11.70$~MHz. Device B~\cite{Eickbusch2022Fast} has a cavity at 5.26~GHz and a transmon
at 6.65~GHz ($-193$~MHz, $T_1=50~\mu$s) with published $E_J=32.33$~GHz, $E_C=181$~MHz and $g=9.12$~MHz, and three measured
facts: no heating of $\ket g$ up to 2000 photons; $\ket e$ ($\ket g$) leaving the computational space above $\sim2500$
($\sim7500$) photons, the first of which its authors reproduced by diagonalization; and $T_1=30~\mu$s at 900 photons, varying from day to
day with narrow features that they attribute to defects.

\begin{figure*}
\includegraphics[width=\textwidth]{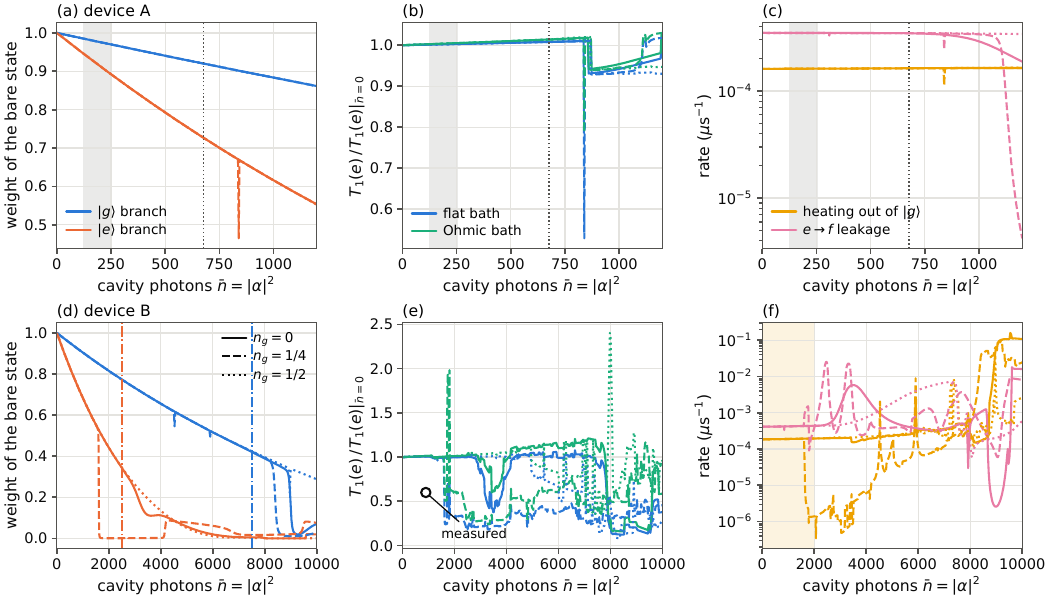}
\caption{GKP devices A (top) and B (bottom) against the cavity photon number, for offset charges $n_g=0$, $\frac14$,
$\frac12$ (line styles). (a), (d) Weight of the bare state in the Floquet branches of $\ket g$ and $\ket e$. (b), (e)
$T_1(e)$ relative to $\nbar=0$ (flat and Ohmic bath). (c), (f) Heating out of $\ket g$ and $e\to f$ leakage. Device A:
shaded, the large ECD displacement; dotted, the DAC limit. Device B, measured: thresholds at $\approx2500$ ($\ket e$) and
$\approx7500$ photons ($\ket g$, dash-dotted), $T_1(e)=30~\mu$s at 900 photons (circle), no heating up to 2000 photons
(shaded).}
\label{fig:gkp}
\end{figure*}

\textit{Floquet--Markov treatment.}---In the frame displaced by the cavity's coherent amplitude, the cavity acts on the
transmon as a classical charge drive of amplitude $\varepsilon=2g\sqrt{\nbar}$ at the cavity
frequency~\cite{Dumas2024MeasurementInduced}. We follow the Floquet states of the Sambe Hamiltonian from $\nbar=0$ by overlap
and compute their rates with the Floquet--Markov golden rule, $\Gamma_{a\to b}=\sum_k|X^k_{ba}|^2\,\Gamma(q_a-q_b-k\omega_d)$
with $X^k_{ba}=\sum_m\langle\phi_{b,m}|\hat n|\phi_{a,m+k}\rangle$ and quasienergies $q$, for a flat charge bath (Ohmic as a
sensitivity case) referenced to the bare $T_1$, with detailed balance at the measured thermal population. The generalized
parity of the driven transmon (charge reflection combined with a half-period shift) makes $X^k$ vanish between states of
equal parity for even $k$ and of opposite parity for odd $k$~\cite{Cohen2023Reminiscence}. The treatment reproduces the
undriven energies and rates, is independent of the replica chosen for the final state, matches exact diagonalization (cavity
basis up to 300 photons) in the $e$--$g$ Stark shift at 1--256 photons (0.06\%; 46.6~kHz per photon against the published
46.5~kHz), and its one-photon
$f\to g$ sideband agrees with Eq.~\eqref{eq:M} to 1\% at 133 photons and 5\% at 533 (Table~\ref{tab:law}).

\textit{Device A} (Fig.~\ref{fig:gkp}, top).---$T_1(e)$, decay to $g$ together with the thermal excitation to $f$, stays at
255--257~$\mu$s between 0 and 676 photons for all three
offset charges, and the heating out of $\ket g$ ($1.61\times10^{-4}/\mu$s) and the thermal $e\to f$ ($3.47\times10^{-4}/\mu$s)
do not depend on $\nbar$; an $f$ level would acquire $f\to g$ at $7.1\times10^{-6}/\mu$s at 256 photons ($2.0\times10^{-5}$ at
676), against $f\to e$ at $6.8\times10^{-3}/\mu$s. There is no ionization up to 1200 photons, apart from a narrow avoided
crossing of the $e$ branch at 839 photons for $n_g=\frac14$ (cf.\
Refs.~\cite{Shillito2022Dynamics,Khezri2023MeasurementInduced,Dumas2024MeasurementInduced}). The measured leakage of
$6.8\times10^{-4}$ per round matches thermal $e\to f$ ($\approx8.5\times10^{-4}$ per half-cycle with $P(e)\approx\frac12$),
so the half of the logical bit-flip error that the authors' budget leaves unexplained is not smooth photon-induced ancilla
decay.

\textit{Device B} (Fig.~\ref{fig:gkp}, bottom).---The model reproduces two of the three measured facts. The heating out of
$\ket g$ stays thermal ($1.89$ and $1.96\times10^{-4}/\mu$s at 900 and 2000 photons), and $\ket e$ ionizes sharply at 1610
photons for $n_g=\frac14$ and gradually between $\sim1700$ and 4300 photons at the sweet spots, bracketing the $\sim2500$
measured over a drifting offset charge; the $\ket g$ branch weight falls below $\frac14$ at 8400--8960 photons against
$\sim7500$ measured. The third fact is not reproduced: $T_1(e)=49.4~\mu$s at 900 photons against $49.0~\mu$s bare, for all
offset charges, so the measured $30~\mu$s is not smooth photon-induced decay, consistent with the authors' attribution to
defects.

\section{Logical memory and the public raw data}
\label{app:logical}

\textit{Logical bit flips and erasures.}---Fitting $\langle Z_L(0)Z_L(n)\rangle$ per initial state in the public raw
data~\cite{Putterman2024Data}, as in Ref.~\cite{Putterman2025Hardwareefficient}, reproduces the published logical
bit-flip error per cycle (its Fig.~4b) to 0.1--0.6\% (statistical errors 1.4--2.1\%). The measured single-gate rates of
its Fig.~16 plus the measured idle rates account for the logical bit-flip rate of the distance-5 code within 10\% (of the
distance-3 sections within 17\%), so no mechanism specific to the repetition code is needed, as stated in
Ref.~\cite{Putterman2025Hardwareefficient}. Our model, $\varepsilon_L=\frac12[1-\prod_s(1-2L_s)\prod_j(1-2p_{{\rm
idle},j})]$ with the measured idle rates $p_{{\rm idle},j}$, is 1.4--5.4 times below the data with the idle-referenced bath
and within 0.95--1.42 of them with $\lambda=3$ (Sec.~\ref{sec:ocelot}), for all three code sections at $|\alpha|^2=2$--4;
for the gate part alone (data minus measured idle rates) the factors are 3.8--7.2 and 0.9--1.4. This carries the
gate-level factor to the logical level and does not confirm it independently. After unmixing the syndrome changes between
rounds (10--36\% per round, mostly phase flips), the erasure probability per round on the $g$ side is 4.6--6.6\%, with
slopes of only 0.04--0.15\% per photon in the distance-5 run, as $\chi$-matching predicts; the
readout side adds 2.2--5.4\%. Against our model's herald over the ancilla's two
gates and the wait for the other layer, 1.7--3.1\% with the idle-referenced bath, this bounds the on-gate $f\to e$
enhancement by $\lambda_f\le2.6$--3.6 (upper bounds, because assignment errors also count), against the $\lambda\approx3$
that the gate and logical data require. There are no same-round erasure correlations between ancillas and little
persistence between rounds.

\textit{Logical phase flips and measurement errors.}---Stim circuits~\cite{Gidney2021Stim} of the phase-flip memory (cat $Z$
errors in the two windows of each cycle, heralded erasures, measurement errors), decoded by correlated
matching~\cite{Higgott2025Sparse} with weights from a data-anchored model, reproduce the published logical phase-flip rates
from the public syndromes (Fig.~3f of Ref.~\cite{Putterman2025Hardwareefficient}; e.g.\ $0.44\%$, $2.10\%$ and $7.33\%$ per
cycle for $d=5$ at $|\alpha|^2=1$, 2 and 3.5). Our model with the same structure is within $-5\ldots+12\%$ of the decoded data
when anchored on the measured lifetimes and error rates, and 20--43\% low with idle-referenced inputs, mainly because the
storage decays 1.1--1.4 times faster during the gate period than at idle (Fig.~30 of
Ref.~\cite{Putterman2025Hardwareefficient}) and, at $|\alpha|^2\ge2$, because the measured syndrome error grows. The
syndrome error per round, from the time-like detector pairs, grows from
1.2--2.0\% at $|\alpha|^2=1$ to 5.0, 5.4, 1.7 and 12.9\% at $|\alpha|^2=3.5$ for the four ancillas; the experiment attributes
this to conditional storage Kerr terms with the gate length fixed~\cite{Putterman2025Hardwareefficient}. In our dressed ladders
the branch-phase curvature ($-2$ to 17~mrad per photon$^2$) explains 7--47\% of this growth, largest for A4 and smallest for
A3 as in the data; the rest
would need a linear phase error of 0.1--0.2~rad per photon, which the bit flips barely register at $|\alpha|^2\ge3$.

\textit{As built.}---The overall logical error $\varepsilon_L=(\varepsilon_{\rm bit}+\varepsilon_{\rm phase})/2$ of the
idle-referenced model has its minimum within $-17\%\ldots0\%$ of the data and $\nbar^*$ within 0.5 photons, partly by
compensation: near the optimum, where the idle part dominates, the bit flips are up to about 40\% high (in the distance-5
code 10--20\%), the gate bit flips low at $|\alpha|^2\ge2.5$ and the phase flips 20--40\% low.

\section{Heralded or silent}
\label{app:herald}

The relaxation of a rotated cat emits a burst of photons through the buffer, and monitoring it could turn bit flips into
erasures~\cite{Ferrari2026Bit}. We compute the exact click statistics of the re-stabilization: the no-click generator
$\mathcal L-\eta\mathcal J$ (detector efficiency $\eta$, jump $a^2-\nbar$) has the same dark code space, and the $n$-click
distribution follows from $v_{n+1}=\eta\mathcal J(-\mathcal L_\eta)^{-1}(1-\mathcal P_\eta)v_n$. A decay during a gate leaves
a partial rotation that relaxes with a burst, but a gate skipped because the ancilla decayed to $g$ before the second cat's
CX leaves that cat in the code space, silent at any efficiency; and every gate makes the normal re-stabilization emit
through the deterministic shear, so that a single-photon threshold at $\eta=0.6$ gives false heralds 20--300 times the true
error rate. With a threshold that keeps false heralds below the error rate, 31--39\% of the logical-relevant errors are
heralded at $\eta=0.6$ with $\lambda=3$ at $|\alpha|^2=4$--8 (17--21\% with the idle-referenced bath at $|\alpha|^2=4$), and
at most about half with an ideal detector.

\section{Switching the stabilization}
\label{app:switch}

In Ocelot the two-photon stabilization is off during the gates and on for the readout and reset of the ancillas; this
adds no bit flips of its own in our models. In an exact single-mode cycle map, pulsing with the pumped coherence kept
during the gate is neutral, $p({\rm pulsed})/p({\rm static})=0.96$--$1.04$ at
$|\alpha|^2\le1.5$. With the buffer as a mode, the switch-on transient of an ideal cat is $1$--$4\times10^{-8}$ at
$|\alpha|^2=2$--4 for pump and drive ramped together in 30~ns, because they enter as
$g_2s(t)[(a^2-\nbar)b^\dagger+{\rm h.c.}]$, which annihilates the code states; only a pump ahead of the drive costs, 3.5--12
times the idle rate (S1, $|\alpha|^2=4$). The same window model reproduces the measured cost of pulsed stabilization of the
companion device~\cite{Putterman2025Preserving} at $|\alpha|^2\le3$ without a fit. Idle noise during the
$\sim1/(2\kappa_2\nbar)$ relaxation after a gate raises the conversion of moderately rotated states (0.5~rad) by 17--44\%;
the fault tables use the noise-free map, which therefore underestimates the logical-relevant error by up to 15\% at
$|\alpha|^2=2$ and 6\% at 4, and negligibly at $|\alpha|^2\ge8$ or with $\lambda=3$.

\section{Defects swept through the qubit lines and during the flux pulses}
\label{app:stark}

The pulsed $T_1$ of A4 (Fig.~37d of Ref.~\cite{Putterman2025Hardwareefficient}) shortens for $e$ with $|\alpha|^2=4$ in the
storage ($30.3\to23.9~\mu$s), while $T_1(f)$ changes by only $6\pm6\%$ ($16.0\to15.1~\mu$s). The $e\to g$ line shifts by
$\chi_{ge}$ per photon ($-1.0$ to $-2.7$~MHz in all pairs), whereas $\chi$-matching keeps the $f\to e$ line nearly fixed
($+0.06$, $+0.18$~MHz at $n=1$, 4 for A4--S5), so a frequency-selective bath swept by the photon number acts on $T_1(e)$ and
not on $T_1(f)$, as when a qubit is Stark-shifted into defects during
readout~\cite{Carroll2022Dynamics,Thorbeck2024Readoutinduced,Huang2026Readoutinduced}. In the literature ensembles of
App.~\ref{app:tls}, one defect within a few MHz of the swept $e\to g$ line reproduces this pattern (decay rates during the
on-periods raised at least 1.5 times for $e$ and at most 1.3 times for $f$, against $\approx1.8$ and $\approx1.2$
observed) in 4.9--7.0\% of the realizations; the rarer cases in which $f$ is affected come from defects
near the swept opened channel. Charge-coupled Markovian channels are excluded by parity, and dressed dephasing by white
frequency noise at the measured $T_\varphi$ is 18--35 times too weak. Such a dip of $T_1(e)$ leaves $T_1(f)$ unchanged, sits
near $\nbar\approx\delta_T/|\chi_{ge}|$ ($\delta_T$, the defect's detuning from the idle line), recovers beyond and drifts
between runs, as defects do~\cite{Klimov2018Fluctuations}. Landau--Zener passages through defects during the flux pulses do
not explain the pulsed-versus-held-on difference of Fig.~37d either: over 20\,000 idle-conditioned realizations the mean
extra decay per pulse is $\approx2\times10^{-4}$, and an extra decay of the size that difference needs occurs in
0.08--0.15\% of them for $e$ and 0.01--0.02\% for $f$.

\section{The logical floor against the distance}
\label{app:floor}

For storage lifetimes of 300~$\mu$s and beyond, the minimum logical error of the projection of App.~\ref{app:projection}
rises with $d$ beyond $d\approx3$--9 [Fig.~\ref{fig:projection}(a) and Table~\ref{tab:projection}]: the bit-flip part
grows linearly with $d$ while the phase part is already small, so the transmon-ancilla CX sets a logical floor that a
storage $T_1$ beyond about 1~ms hardly lowers, for a constant floor as well as for ours. Estimates of logical errors of $10^{-5}$ and
below for this architecture~\cite{Putterman2025Hardwareefficient} assume idealized gates; with transmon ancillas the CX bit
flips per stabilizer must first drop by two to three orders of magnitude, for instance with cat--cat
gates~\cite{Ye2026Biaspreserving}.

\end{document}